\documentclass[11pt]{article}
\usepackage[utf8]{inputenc}
\usepackage[T1]{fontenc}
\usepackage{lmodern}
\usepackage[a4paper,margin=1in]{geometry}
\usepackage{amsmath,amssymb,amsfonts}
\usepackage{graphicx}
\usepackage{booktabs,longtable,array,calc}
\usepackage{setspace}
\usepackage{hyperref}
\usepackage{caption}
\usepackage{enumitem}
\usepackage{float}
\usepackage{microtype}
\title{Measuring Human Behavior Through Controlled Perturbations: \\
A Framework for Behavioral System Identification}
\author{Pietro Cipresso\\Department of Psychology, University of Turin\\\texttt{pietro.cipresso@unito.it}}
\date{}
\begin{document}
\maketitle
\begin{abstract}
Human behavior is typically measured by relating observed responses to
latent traits inferred from statistical models. Although this approach
has produced powerful tools for psychological assessment, it largely
treats behavior as a static property of individuals rather than as the
observable output of an evolving dynamical system. Yet most behavioral
phenomena emerge from continuous interactions between internal states
and environmental conditions, generating trajectories of actions,
decisions, and physiological responses over time.

Here we propose a framework in which human behavior is measured by
reconstructing the dynamical systems that generate it through controlled
perturbations. In this perspective, experimental environments act as
measurement instruments that probe behavioral dynamics through
structured perturbations. Behavioral observations are interpreted as the
outputs of stochastic controlled dynamical systems, and behavioral
measurement becomes a system identification problem: generative models
are reconstructed so as to reproduce the trajectory distributions
produced by human behavior under equivalent perturbation conditions.

We formalize this framework using stochastic dynamical systems,
introduce criteria for reconstruction and functional equivalence under
unseen perturbations, and analyze the roles of perturbation design,
identifiability, and experimental domain boundaries. The approach is
illustrated across multiple behavioral domains, including affective
dynamics, cognitive flexibility, stress reactivity, and social
interaction.

By integrating controlled perturbations, dynamical modeling, and
trajectory-level validation, this framework reframes behavioral
measurement as the reconstruction of generative behavioral dynamics.
More broadly, it suggests a path toward a metrology of human behavior in
which programmable experimental environments, such as immersive virtual
reality platforms, serve as instruments capable of systematically
probing and characterizing behavioral systems.
\end{abstract}

\section{Introduction: The Measurement Problem in Behavioral Science}

Human behavior is widely studied across the behavioral sciences, yet the
question of how behavior should be measured remains largely unresolved
(Borsboom, D., 2005). From psychology and neuroscience to economics and
human--computer interaction, researchers seek to identify the processes
that generate observable behavioral patterns, an objective that has also
been addressed in probabilistic and generative frameworks such as the
free-energy principle (Friston, K., 2010). Yet despite this common goal,
most measurement frameworks primarily focus on internal generative
models and treat behavior as a static property of individuals rather
than as the observable output of a dynamical system evolving over time,
an idea that has long been emphasized in dynamical approaches to
cognition and behavior (Molenaar, P. C. M., 2004; van Gelder, T., 1995;
Thelen, E., \& Smith, L. B., 1994).

Human behavior unfolds through continuous interactions (Kelso, J. S.,
1995) between internal states and environmental conditions, giving rise
to temporally structured patterns that have been extensively studied
within dynamical systems frameworks (Kelso, J. S., 1995). Capturing
these dynamics requires measurement approaches that move beyond static
latent traits toward the reconstruction of the systems that generate
behavioral trajectories (Hamaker, E. L., \& Wichers, M., 2017; Hamaker
EL, Asparouhov T, Brose A, Schmiedek F, Muthén B, 2018; Kuppens, P.,
Oravecz, Z., \& Tuerlinckx, F., 2010; Speyer, L. G., Murray, A. L., \&
Kievit, R., 2024).

Early psychometric theory already recognized the importance of
establishing a rigorous measurement framework for behavioral phenomena.
In particular, Rasch emphasized that measurement in the behavioral
sciences should aim to establish invariant relations between
observations and latent variables, independent of the particular sample
of items or persons used in a study.

One reason for this discrepancy lies in a longstanding methodological
constraint in behavioral research. Experimental paradigms have
historically faced a trade-off between ecological validity and
experimental control (Bronfenbrenner, U., 1977; Pooja R, Ghosh P,
Sreekumar V., 2024). Laboratory experiments provide high levels of
control and repeatability but typically rely on simplified tasks that
capture only limited aspects of real-world behavior. In contrast,
naturalistic observation and experience sampling methods allow
researchers to study behavior in realistic contexts but provide little
control over the perturbations experienced by participants (Shiffman S,
Stone AA, Hufford MR, 2008; Csikszentmihalyi, M., Larson, R., 2014;
Kahneman, D., Krueger, A. B., Schkade, D. A., Schwarz, N., \& Stone, A.
A., 2004; Fritz, J., Piccirillo, M. L., Cohen, Z. D., Frumkin, M.,
Kirtley, O., Moeller, J., Neubauer, A. B., Norris, L. A., Schuurman, N.
K., Snippe, E., \& Bringmann, L. F., 2024; Xie, K., Vongkulluksn, V. W.,
Heddy, B. C., \& Jiang, Z., 2024). As a result, behavioral measurement
has often relied on statistical inference from observational data rather
than on experimental probing of behavioral systems.

Recent technological developments are beginning to alter this landscape.
Advances in immersive experimental environments, multimodal sensing, and
computational modeling make it increasingly feasible to observe rich
behavioral trajectories while maintaining experimental control over
environmental conditions (Bohil CJ, Alicea B, Biocca FA, 2011; Cipresso
P, 2015; Slater M and Sanchez-Vives MV, 2016; Faria AL, Latorre J, Silva
Cameirão M, Bermúdez i Badia S and Llorens R, 2023; Quirós-Ramírez MA,
Feineisen A, Streuber S, Reips U-D, 2025). In particular, immersive
environments, such as those enabled by virtual reality technologies,
allow researchers to construct programmable experimental contexts in
which environmental perturbations can be systematically manipulated
while preserving a high degree of experiential realism. At the same
time, modern sensing technologies allow the continuous recording of
behavioral signals such as movement, gaze, physiological responses, and
interaction patterns (Bohil CJ, Alicea B, Biocca FA, 2011; Potter LN,
Yap J, Dempsey W, Wetter DW, Nahum-Shani I, 2023).

These developments open the possibility of approaching behavioral
measurement from a different perspective. Instead of treating behavioral
observations as isolated responses from which latent traits are
inferred, behavioral systems can be studied by systematically probing
them with controlled perturbations and analyzing the resulting
behavioral trajectories. This approach closely resembles the logic of
system identification in engineering and control theory, where the goal
is to reconstruct system dynamics from input--output data through
carefully designed inputs (Ljung L., 2010; Bellman, R., \& Åström, K.
J., 1970; Söderström, T., \& Stoica, P., 1989).

The framework developed here aims to provide a conceptual foundation for
a more dynamical and experimentally grounded approach to behavioral
measurement. We first formalize behavioral systems as stochastic
dynamical processes interacting with controlled perturbations. We then
introduce criteria for reconstructing generative models from
perturbation--response trajectories and for evaluating functional
equivalence between reconstructed models and behavioral systems.
Finally, we discuss how this perspective relates to existing traditions
in psychometrics and computational modeling and how it may contribute to
the development of a metrology of human behavior (Cipresso P. and
Immekus J.C., 2017).

In this work, we introduce a perturbation-based framework for reconstructing behavioral systems from trajectory data generated under controlled environmental perturbations. By integrating controlled perturbations, dynamical modeling, and trajectory-level validation, the proposed approach reframes behavioral measurement as a system identification problem. We formalize the framework, illustrate its implementation across behavioral domains, and discuss its implications for psychometrics, computational modeling, and the development of a metrology of human behavior.

\section{The Limits of Current Measurement Paradigms}

Despite substantial progress in the behavioral sciences, the problem of
how to measure human behavior remains only partially resolved. Existing
approaches have produced powerful tools for quantifying individual
differences and for modeling observed responses, yet they are
constrained by fundamental methodological limitations that restrict
their ability to capture the dynamical nature of behavior.

A central feature of most traditional measurement frameworks is the
reliance on latent variables inferred from patterns of observed
responses. In psychometrics, models such as factor analysis and item
response theory establish probabilistic relationships between latent
traits and observable outcomes, enabling the estimation of individual
differences along relatively stable dimensions (Borsboom, D., 2005;
Cronbach, L. J., \& Meehl, P. E., 1955). While these approaches have
achieved high levels of reliability and scalability, they implicitly
treat behavior as a static or slowly varying property of individuals.
Temporal structure is often reduced to aggregated scores, and the
underlying processes that generate behavioral trajectories remain
largely unmodeled.

Parallel developments in ecological and ambulatory assessment have
attempted to overcome this limitation by capturing behavior in
naturalistic contexts. Experience sampling methods and related
approaches provide access to behavior as it unfolds in daily life,
increasing ecological validity and enabling the study of within-person
variability (Kahneman, D., Krueger, A. B., Schkade, D. A., Schwarz, N.,
\& Stone, A. A., 2004; Shiffman S, Stone AA, Hufford MR, 2008; Fritz,
J., Piccirillo, M. L., Cohen, Z. D., Frumkin, M., Kirtley, O., Moeller,
J., Neubauer, A. B., Norris, L. A., Schuurman, N. K., Snippe, E., \&
Bringmann, L. F., 2024). However, these methods introduce a different
limitation: the absence of experimental control over the perturbations
experienced by individuals. Because environmental inputs are neither
controlled nor systematically varied, it becomes difficult to
disentangle the causal structure underlying observed behavioral
trajectories. As a result, inference remains largely observational, and
the identification of generative mechanisms is limited.

Laboratory-based experimental paradigms occupy an intermediate position,
offering precise control over stimuli and high levels of repeatability.
Controlled tasks allow researchers to manipulate specific variables and
test well-defined hypotheses, often with fine temporal resolution. Yet
this control is typically achieved at the cost of ecological validity.
Simplified tasks capture only restricted aspects of real-world behavior,
and the resulting measurements may fail to generalize beyond the
experimental setting. Consequently, laboratory paradigms often provide
high internal validity but limited insight into the broader dynamical
structure of behavior in natural environments (Bohil CJ, Alicea B,
Biocca FA., 2011 ; Bronfenbrenner, U., 1977; Cipresso, P., 2015; Faria
AL, Latorre J, Silva Cameirão M, Bermúdez i Badia S and Llorens R,
2023).

These three approaches reflect a longstanding trade-off in behavioral
research between ecological validity and experimental control.
Naturalistic methods capture realistic behavior but lack
controllability, while laboratory experiments provide control but at the
expense of realism. Traditional psychometric models, in turn, offer
scalable measurement but abstract away from temporal dynamics and
environmental interaction. None of these paradigms simultaneously
provides controlled, repeatable, and ecologically meaningful
observations of behavioral trajectories (Shiffman S, Stone AA, Hufford
MR, 2008; Bronfenbrenner, U., 1977).

This limitation has important consequences for the development of a
rigorous measurement science of behavior. Without controlled
perturbations, it is difficult to treat behavioral systems as objects
that can be systematically probed and identified (Åström, K. J., 1995;
Åström, K. J., \& Eykhoff, P., 1971). Without temporally rich
observations, it is difficult to characterize the dynamical processes
underlying behavioral change. Without reproducible experimental
conditions, it is difficult to establish invariance and comparability
across studies.

The result is that behavioral measurement remains largely indirect.
Rather than identifying the mechanisms that generate behavior, current
approaches typically infer latent structures from observed data or
describe statistical regularities in behavioral patterns. While these
strategies have proven useful, they do not fully address the challenge
of reconstructing the generative systems that produce behavior.

These limitations motivate the need for a different measurement
paradigm. A framework that integrates controlled perturbations,
temporally extended observations, and reproducible experimental
environments would make it possible to treat behavioral systems as
dynamical entities that can be systematically interrogated. Such a
framework would move beyond the estimation of static traits and toward
the reconstruction of the processes that generate behavioral
trajectories (Hamaker, E. L., \& Wichers, M., 2017; Molenaar, P. C. M.,
2004; Speyer, L. G., Murray, A. L., \& Kievit, R., 2024).

\section{A Perturbation-Based Perspective on Behavioral Measurement}

The limitations of current measurement paradigms suggest the need for a
shift in how behavioral phenomena are conceptualized and measured.
Rather than treating behavior as a static property to be inferred from
observed responses, it may be more appropriate to view behavior as the
observable output of a dynamical system interacting with its
environment. Within this perspective, measurement is not a passive
process of observation, but an active process of system interrogation
(Ljung, 2010).

In many areas of science, the identification of dynamical systems relies
on the application of controlled inputs, whose design critically
determines the informativeness of the resulting data (Gevers, M., \&
Ljung, L., 1986). This principle is central to system identification,
where the properties of an unknown system are inferred by probing it
with structured perturbations. Extending this idea to behavioral science
suggests that meaningful measurement requires not only observing
behavior, but actively shaping the conditions under which behavior
unfolds.

From this viewpoint, environmental conditions can be interpreted as
inputs to a behavioral system, while observable actions, decisions, and
physiological responses constitute its outputs. Behavioral trajectories
emerge from the continuous interaction between these inputs and the
internal dynamics of the system. Measurement, therefore, becomes the
problem of reconstructing the system that generates these trajectories,
rather than summarizing the trajectories themselves. A key implication
of this perspective is that perturbations are not merely experimental
manipulations introduced to test specific hypotheses, but constitute the
core mechanism through which behavioral systems can be identified.
Carefully designed perturbations determine the informativeness of the
resulting data, enabling the experimenter to explore the space of
possible system responses and reveal aspects of the underlying dynamics
that would remain unobservable under passive observation. The
informativeness of measurement is thus directly linked to the structure
and diversity of the perturbations applied.

This shift also redefines the role of experimental environments. Instead
of serving solely as contexts in which behavior is observed,
environments become measurement instruments that implement controlled
perturbations. Advances in programmable and immersive environments,
including VR-based systems, make it increasingly feasible to construct
experimental settings that are both ecologically meaningful and
precisely controllable. These environments enable the systematic
exploration of behavioral dynamics under reproducible conditions (Bohil
CJ, Alicea B, Biocca FA, 2011; Slater M and Sanchez-Vives MV, 2016;
Cipresso, P., 2015; Cipresso, P., Giglioli, I. A. C., Raya, M. A., \&
Riva, G., 2018; Chirico, A., Yaden, D. B., Riva, G., \& Gaggioli, A.,
2016; Riva, G., 2009; Riva, G., Wiederhold, B. K., \& Mantovani, F.,
2019).

Within this framework, the goal of behavioral measurement is to
reconstruct a model of the system that is capable of reproducing the
observed trajectories under a range of perturbations. Importantly, the
validity of such a model cannot be established solely by its ability to
fit observed data, but must be evaluated based on its capacity to
generate behavior that is indistinguishable from that of the original
system when exposed to new perturbations. This notion of functional
equivalence shifts the focus from descriptive adequacy to generative
fidelity (Box, 1976).

Adopting a perturbation-based perspective thus transforms behavioral
measurement into a problem of dynamical system reconstruction under
controlled experimental conditions. This perspective provides a unifying
framework that integrates experimental design, data collection, and
modeling into a single process, and sets the stage for a more principled
and reproducible science of human behavior.

\section{Formal Framework for Perturbation-Based Behavioral
Reconstruction}\label{formal-framework-for-perturbation-based-behavioral-reconstruction}

The central premise of the present framework is that human behavior
should be understood not merely as a set of observable responses but as
the manifestation of an underlying dynamical system interacting with
structured environmental perturbations. In this perspective,
psychological measurement is reframed from the estimation of latent
scores to the reconstruction of generative behavioral dynamics. The
objective is therefore not simply to predict responses but to identify a
model capable of reproducing the trajectories produced by a behavioral
system when subjected to controlled interventions (Schwarz, G., 1978;
Borel, A., 2012).

The overall structure of the perturbation-based behavioral
reconstruction framework is summarized in Figure 1. The framework
conceptualizes behavioral measurement as a system identification process
in which controlled perturbations probe latent behavioral dynamics,
generating trajectory data used to reconstruct and validate generative
models of behavior.

\begin{figure}[htbp]
\centering
\includegraphics[width=\textwidth]{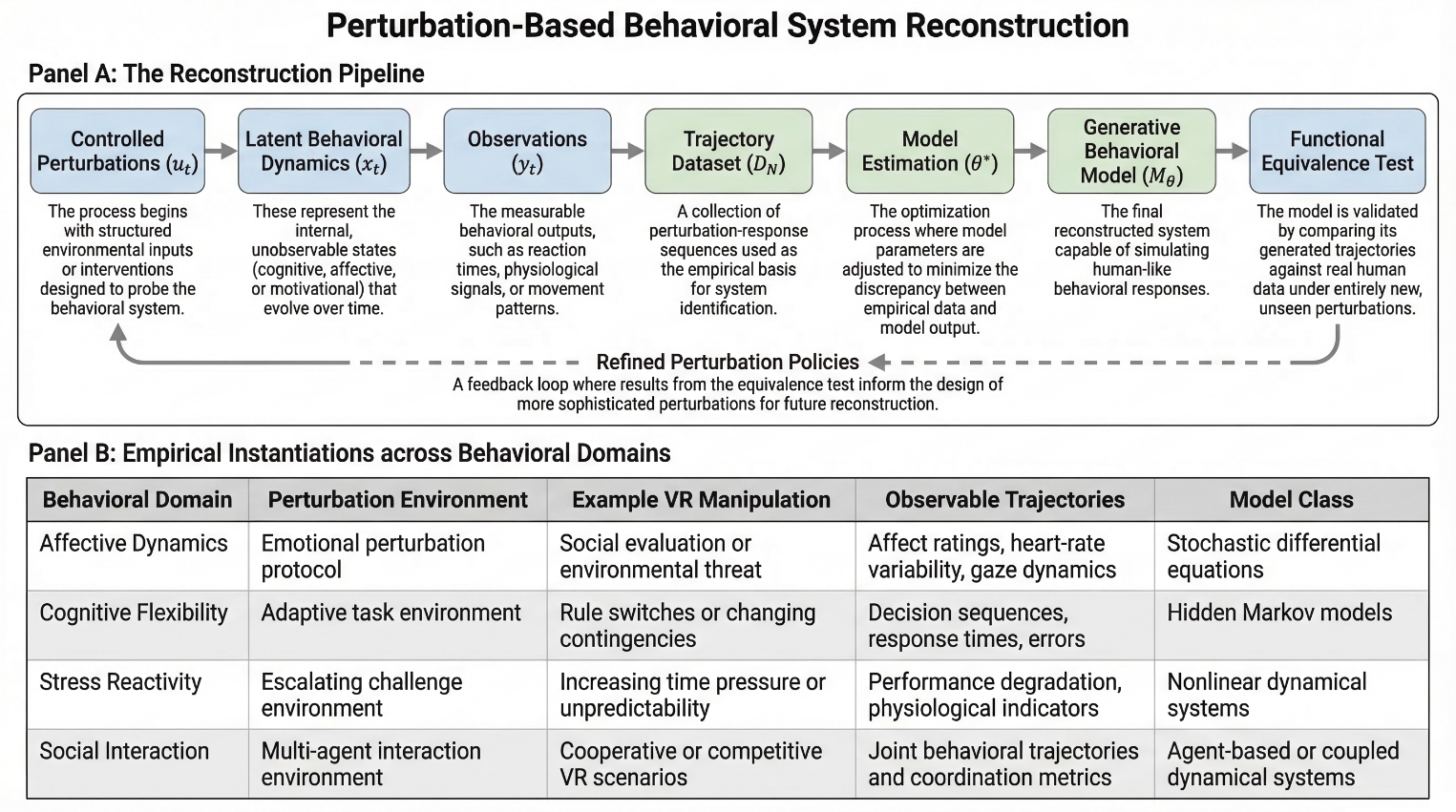}
\caption{Perturbation-based behavioral system reconstruction framework. Behavioral measurement is formulated as a system identification problem in which controlled perturbations $u_t$ probe latent behavioral states $x_t$, producing observable trajectories $y_t$. These perturbation--response trajectories form a dataset $D_N$ used to estimate a generative behavioral model $\mathcal{M}_{\theta}$. Reconstruction is validated through functional equivalence tests evaluating whether the model generates trajectory distributions indistinguishable from those produced by the human system under previously unseen perturbations. The validation process informs the design of refined perturbation policies, forming an iterative experimental loop. The lower panel illustrates representative behavioral domains and corresponding model classes that can instantiate the reconstruction framework.}
\end{figure}

To formalize this idea, we represent a behavioral system as a stochastic
controlled dynamical system (Oksendal, B., 2013; Pavliotis, G. A.,
2014). Let \(x_{t}\) denote the latent behavioral state of an individual
at time \(t\) , \(u_{t}\) a perturbation applied by the experimental
environment, and \(y_{t}\) the observable behavioral output. The
evolution of the system is described by the pair of equations

\[x_{t + 1} \sim Q( \cdot \mid x_{t},u_{t}),\quad\quad y_{t} \sim R( \cdot \mid x_{t}),\]

where \(Q\) is a transition kernel describing how the internal state
evolves under perturbations and \(R\) is an observation kernel mapping
latent states to measurable outputs. The state \(x_{t}\) may represent
cognitive, affective, or motivational variables that are not directly
observable, whereas the output \(y_{t}\) may correspond to reaction
times, movement trajectories, physiological signals, or affective
reports.

Given a sequence of perturbations
\(u_{0:T - 1} = (u_{0},\ldots,u_{T - 1})\) , the behavioral system
induces a probability distribution over observable trajectories
\(y_{0:T} = (y_{0},\ldots,y_{T})\) . We denote this trajectory law by

\[P_{\mathcal{B}}(y_{0:T} \mid u_{0:T - 1}),\]

where \(\mathcal{B}\) refers to the underlying behavioral system. This
distribution constitutes the fundamental object of measurement within
the present framework. Unlike classical psychometric approaches that
summarize behavior through static scores or averaged responses, the
trajectory law captures the full temporal evolution of behavior under
controlled perturbation.

As illustrated in Figure 1, the perturbation sequence \(u_{t}\) acts as
an external input probing the latent behavioral state \(x_{t}\), while
the observable outputs \(y_{t}\) correspond to measurable behavioral
trajectories generated by the underlying dynamical system.

A central but often implicit assumption in behavioral modeling concerns
the unit of reconstruction. In the present framework, two distinct
formulations must be distinguished. In the idiographic formulation, the
goal is to reconstruct the dynamics of a single behavioral system
\(B^{(i)}\), corresponding to an individual. The dataset then consists
of multiple perturbation-response trajectories collected from the same
system under varying conditions:

\[D_{N} = \{(u_{0:T_{k} - 1}^{(k)},y_{0:T_{k}}^{(k)})\}_{k = 1}^{N}\]

In this setting, reconstruction corresponds to identifying a generative
model that approximates the individual-specific trajectory law
\(P_{B^{(i)}}(y_{0:T} \mid u_{0:T - 1})\). This formulation implicitly
requires assumptions about the stability or ergodicity of the underlying
system across trials.

In contrast, a nomothetic formulation considers a population of
behavioral systems, where each trajectory may originate from a different
individual. In this case, the objective is to reconstruct a
population-level trajectory distribution:

\[P(y_{0:T} \mid u_{0:T - 1}) = \int P_{B}(y_{0:T} \mid u_{0:T - 1})\text{ }dP(B)\]

While this formulation aligns more closely with classical psychometric
approaches, it fundamentally alters the reconstruction problem and may
obscure individual-specific dynamics. As shown by Molenaar (2004), such
aggregation is not generally valid under non-ergodic conditions.

The present framework is primarily grounded in the idiographic
formulation, where controlled perturbations are used to probe and
reconstruct individual behavioral systems. Extensions to
population-level modeling require additional assumptions and are left
for future work.

Empirical behavioral data consist of a set of perturbation-response
trajectories obtained across experimental trials or participants,
depending on the unit of reconstruction adopted. Let

\[\mathcal{D}_{N} = \left\{ (u_{0:T_{i} - 1}^{(i)},y_{0:T_{i}}^{(i)}) \right\}_{i = 1}^{N}\]

denote a dataset containing \(N\) such trajectories. The goal of
behavioral reconstruction is to identify a generative model capable of
reproducing the trajectory law induced by the true system.

\textbf{To illustrate the operationalization of the proposed framework},
consider a cognitive flexibility task implemented in a VR environment in
which participants are exposed to rule-switching perturbations. At each
time step, the perturbation \(u_{t}\) corresponds to changes in task
contingencies, while the observable output \(y_{t}\) consists of action
choices and reaction times. A candidate model \(M_{\theta}\), such as a
switching state-space model, is fitted to a set of perturbation-response
trajectories by minimizing the discrepancy between empirical and
model-generated trajectory distributions. Functional equivalence is then
evaluated by exposing both the participant and the reconstructed model
to novel perturbation sequences not used during estimation and comparing
the resulting trajectory distributions. This example illustrates how
controlled perturbations, trajectory-level data, and generative modeling
are integrated within a concrete experimental setting.

To this end, we consider a family of candidate models

\[\mathfrak{M = \{}\mathcal{M}_{\theta}:\theta \in \Theta\},\]

where each model \(\mathcal{M}_{\theta}\) is itself a stochastic
controlled dynamical system parameterized by \(\theta\) . Reconstruction
consists in estimating the parameter value \(\theta\) that best
reproduces the empirical perturbation--response data. Formally, this can
be expressed as the optimization problem

\[\theta^{\star} = arg\min_{\theta \in \Theta}\mathcal{L}_{N}(\theta),\]

where the loss function

\[\mathcal{L}_{N}(\theta) = \frac{1}{N}\sum_{i = 1}^{N}D\left( P_{\text{emp}}^{(i)},P_{\mathcal{M}_{\theta}}( \cdot \mid u^{(i)}) \right) + \lambda\Omega(\theta)\]

quantifies the discrepancy between empirical trajectory distributions
and those generated by the model. The function \(D\) measures the
distance between trajectory distributions, while \(\Omega(\theta)\)
imposes structural constraints or regularization on the model parameters
(Cover, T. M., \& Thomas, J. A., 2012).

Crucially, reconstruction claims must be evaluated within explicitly
defined experimental boundaries. Behavioral systems are embedded in
complex environments, and it is neither meaningful nor feasible to
reconstruct their dynamics universally. Instead, reconstruction is
defined relative to a bounded experimental domain

\[\mathcal{D = (}\mathcal{U}_{\mathcal{D}},\mathcal{Y}_{\mathcal{D}},T_{\mathcal{D}},\Pi_{\mathcal{D}}),\]

where \(\mathcal{U}_{\mathcal{D}}\) specifies the admissible
perturbations, \(\mathcal{Y}_{\mathcal{D}}\) the admissible output
space, \(T_{\mathcal{D}}\) the temporal horizon, and
\(\Pi_{\mathcal{D}}\) the class of perturbation policies allowed within
the experiment. All statements about reconstruction or equivalence are
therefore restricted to the domain \(\mathcal{D}\) .

Within this bounded domain, the central question becomes whether a
reconstructed model can reproduce the behavior of the true system under
perturbations that were not used during estimation. Let
\(\Pi_{\mathcal{D}}^{train}\) denote the perturbation sequences used
during reconstruction and \(\Pi_{\mathcal{D}}^{test}\) a disjoint set of
perturbations reserved for validation. A reconstructed model
\(\mathcal{M}_{\widehat{\theta}}\) is said to be functionally equivalent
to the behavioral system if the trajectory distributions generated by
the two systems remain sufficiently close under all unseen perturbations
in the test set. Formally, this condition can be written as

\[\sup_{u \in \Pi_{\mathcal{D}}^{test}}D\left( P_{\mathcal{B}}( \cdot \mid u),P_{\mathcal{M}_{\widehat{\theta}}}( \cdot \mid u) \right) \leq \delta,\]

where \(\delta\) represents an admissible tolerance level. Functional
equivalence thus requires that the reconstructed model reproduce the
behavioral dynamics of the true system across novel perturbations rather
than merely fitting the trajectories observed during training.

The discrepancy measure \(D\) may be defined in several ways depending
on the nature of the behavioral data. When trajectories are
deterministic or weakly stochastic, a simple Euclidean discrepancy
between trajectories may suffice. In more complex settings involving
stochastic variability and multimodal dynamics, it is preferable to
compare probability distributions over trajectories. In such cases, a
Wasserstein metric (Villani, C., 2009; Santambrogio, F., 2015) or
kernel-based discrepancy such as maximum mean discrepancy provides a
principled way to measure distances between path distributions (Gretton,
A., Borgwardt, K. M., Rasch, M. J., Schölkopf, B., \& Smola, A., 2012;
Muandet, K., Fukumizu, K., Sriperumbudur, B., \& Schölkopf. B., 2017;
Schölkopf, B., \& Smola, A. J., 2002).

However, this definition relies on the comparison of trajectory laws,
which are not directly observable in practice, as empirical data consist
only of finite samples of trajectories collected under specific
perturbations.

Accordingly, functional equivalence must be formulated as a statistical
indistinguishability problem. For a given perturbation \(u\), let
\(\left\{ y^{(i)} \right\}_{i = 1}^{n}\) denote observed trajectories
and \(\left\{ {\overset{\sim}{y}}^{(j)} \right\}_{j = 1}^{m}\)
trajectories generated by the reconstructed model. A discrepancy measure
\(D\) can then be used to compare the two samples:

\[{\overset{\hat{}}{D}}_{n,m}(u) = D(\{ y^{(i)}\}_{i = 1}^{n},\{{\overset{\sim}{y}}^{(j)}\}_{j = 1}^{m})\]

Functional equivalence is then assessed by evaluating whether the
empirical discrepancy remains below a tolerance threshold across a class
of perturbations:

\[\underset{u \in \Pi^{test}}{\sup}{\overset{\hat{}}{D}}_{n,m}(u) \leq \delta + \epsilon_{n,m}\]

Where \(\epsilon_{n,m}\) accounts for statistical uncertainty due to
finite sampling. In this formulation, functional equivalence corresponds
to the inability to distinguish between observed and model-generated
trajectories given finite data, rather than to exact equality of the
underlying trajectory laws.

This formulation naturally leads to the notion of reconstructability. A
behavioral system is said to be reconstructable within a bounded
experimental domain if there exists a model within the candidate class
capable of achieving functional equivalence under unseen perturbations.
Let \(\mathfrak{M}\) denote the chosen model class and \(\delta\) the
admissible tolerance level. The behavioral system \(\mathcal{B}\) is
reconstructable on domain \(\mathcal{D}\) if there exists
\(\theta^{\star} \in \Theta\) such that

\[\sup_{u \in \Pi_{\mathcal{D}}^{test}}D\left( P_{\mathcal{B}}( \cdot \mid u),P_{\mathcal{M}_{\theta^{\star}}}( \cdot \mid u) \right) \leq \delta.\]

The minimal achievable discrepancy

\[\varepsilon^{\star}\mathfrak{(M,}\mathcal{D) =}\inf_{\theta \in \Theta}\sup_{u \in \Pi_{\mathcal{D}}^{test}}D\left( P_{\mathcal{B}}( \cdot \mid u),P_{\mathcal{M}_{\theta}}( \cdot \mid u) \right)\]

defines the intrinsic reconstruction error associated with the chosen
model class and experimental domain. If \(\varepsilon^{\star}\) exceeds
the tolerance level \(\delta\) , the behavioral system cannot be
reconstructed within the specified domain using the given model class.
Importantly, such failures are informative: they reveal structural
limits on reconstructability and guide the design of richer perturbation
environments or more expressive model families.

In this framework, perturbations play a central methodological role.
Rather than serving merely as experimental stimuli, perturbations act as
identification inputs that probe the internal dynamics of the behavioral
system. The informativeness of a perturbation family can therefore be
characterized by its ability to discriminate between alternative models.
Let \(\Pi_{\mathcal{D}}\) denote the admissible perturbation set. Its
discriminatory power can be defined as

\[\Delta(\Pi_{\mathcal{D}}) = \inf_{\theta_{1} \neq \theta_{2}}\sup_{u \in \Pi_{\mathcal{D}}}D\left( P_{\mathcal{M}_{\theta_{1}}}( \cdot \mid u),P_{\mathcal{M}_{\theta_{2}}}( \cdot \mid u) \right).\]

A perturbation family is informative if
\(\Delta(\Pi_{\mathcal{D}}) > 0\) , meaning that distinct models produce
distinguishable trajectory distributions under at least one admissible
perturbation. This definition highlights the importance of experimental
design: reconstruction is possible only if the perturbation environment
generates sufficiently rich behavioral responses to reveal the
underlying dynamics.

Taken together, these elements define a metrological framework for
behavioral science grounded in system identification and controlled
perturbation. Within bounded experimental domains, behavioral
measurement becomes the problem of reconstructing generative dynamical
models whose responses to unseen perturbations remain indistinguishable
from those produced by the human system itself. This perspective extends
the scope of psychometrics from the estimation of latent variables to
the reconstruction of behavioral dynamics and provides a formal basis
for next-generation computational measurement approaches.

\section{Perturbation Environments and Behavioral System
Emulation}\label{perturbation-environments-and-behavioral-system-emulation}

The formal framework introduced above places controlled perturbations at
the center of behavioral reconstruction. The ability to identify
generative behavioral dynamics depends critically on the richness and
controllability of the perturbation space through which the system is
probed. In practical terms, this requirement translates into the need
for experimental environments capable of delivering structured and
reproducible interventions while simultaneously capturing
high-resolution behavioral trajectories.

Traditional psychological experiments typically manipulate a limited set
of discrete stimuli while measuring responses through aggregated
outcomes such as accuracy rates or response times. Although such
paradigms have proven valuable for testing specific hypotheses, they
provide only sparse information about the underlying dynamical structure
of behavior. The reconstruction framework described here instead
requires environments in which perturbations can be continuously
parameterized, temporally structured, and systematically varied in order
to explore the response surface of the behavioral system.

Let \(u_{t}\) denote the perturbation applied to the system at time
\(t\) . In many classical experimental paradigms, \(u_{t}\) belongs to a
small finite set corresponding to categorical stimulus conditions. In
contrast, perturbation-based reconstruction assumes that the
perturbation space \(\mathcal{U}\) may be multidimensional and
continuously parameterized. For example, perturbations may correspond to
variations in environmental complexity, cognitive load, social
interaction dynamics, uncertainty levels, or temporal constraints. The
perturbation sequence \(u_{0:T - 1}\) therefore defines a trajectory
through the perturbation space that elicits a corresponding behavioral
trajectory.

A perturbation environment can thus be viewed as a mapping

\[\Phi:(s_{t},u_{t}) \rightarrow s_{t + 1},\]

where \(s_{t}\) denotes the state of the external environment and
\(u_{t}\) represents an intervention applied to that environment. The
behavioral system interacts with the environment through a
perception--action loop: the environment generates stimuli affecting the
internal behavioral state \(x_{t}\) , while the individual's actions
modify the environment itself. In this sense, the behavioral system and
the environment together form a coupled dynamical system whose
observable trajectories depend jointly on internal dynamics and external
perturbations.

From the standpoint of reconstruction, the environment serves a dual
role. First, it generates the perturbation signals required to probe the
behavioral system. Second, it provides the measurement infrastructure
through which behavioral outputs are observed. Ideally, the environment
should therefore satisfy three properties: controllability of
perturbations, observability of behavioral responses, and
reproducibility of experimental conditions.

Controllability refers to the ability to manipulate perturbations across
a sufficiently rich set of conditions. Observability refers to the
capacity to measure behavioral outputs with sufficient temporal and
spatial resolution. Reproducibility ensures that perturbation sequences
can be repeated across participants or experimental sessions, allowing
the trajectory distributions induced by different perturbations to be
estimated reliably.

Virtual environments are particularly well suited to meet these
requirements. A virtual environment can be regarded as a programmable
perturbation generator in which environmental parameters can be
manipulated in real time while maintaining precise control over
experimental conditions. Because the environment is generated
computationally, perturbations can be defined along multiple continuous
dimensions and dynamically adapted during the interaction.

Formally, let \(\mathcal{U}\) denote the perturbation space and let
\(\Pi_{\mathcal{D}}\) be the set of admissible perturbation policies
defined within the bounded experimental domain introduced earlier. A
virtual environment implements a perturbation policy

\[u_{t} = \pi(s_{t},t),\]

where \(\pi \in \Pi_{\mathcal{D}}\) determines how perturbations evolve
as a function of environmental state and time. This formulation allows
perturbations to depend not only on predetermined experimental schedules
but also on the evolving behavior of the participant. Adaptive
perturbation policies are particularly valuable for system
identification because they can actively explore regions of the state
space that are most informative for distinguishing competing models.

The second critical property of perturbation environments concerns
observability. Modern experimental platforms allow the simultaneous
recording of multiple behavioral signals, including motor trajectories,
gaze patterns, physiological responses, and interaction dynamics. Let

\[y_{t} = (y_{t}^{(1)},y_{t}^{(2)},\ldots,y_{t}^{(k)})\]

denote a multidimensional observation vector capturing these signals.
Each component corresponds to a different behavioral modality, such as
movement velocity, eye fixation coordinates, or physiological indices.
The observation model introduced earlier,

\[y_{t} \sim R( \cdot \mid x_{t}),\]

therefore represents a multimodal measurement process in which latent
behavioral states generate observable signals across multiple sensory
channels.

The integration of multiple observation modalities improves the
identifiability of behavioral dynamics. If distinct latent states
produce indistinguishable outputs in a single modality, additional
modalities may reveal differences that enable model discrimination. From
an information-theoretic perspective, multimodal sensing increases the
effective observability of the latent state space.

A third essential property of perturbation environments is
reproducibility. Because virtual environments are defined
computationally, the same perturbation sequence can be reproduced
exactly across different experimental sessions. This reproducibility
ensures that the trajectory distribution

\[P_{\mathcal{B}}(y_{0:T} \mid u_{0:T - 1})\]

can be estimated consistently across participants exposed to identical
perturbation policies. Reproducibility also allows the separation of
training and testing perturbation sets required for evaluating
functional equivalence. Perturbation sequences used to reconstruct the
model can be strictly separated from those used to validate the
reconstructed system under unseen conditions.

Within this framework, virtual environments act as behavioral system
emulators. They generate controlled perturbations, record behavioral
trajectories, and maintain a fully observable record of environmental
dynamics. The environment therefore provides the experimental substrate
on which reconstruction algorithms operate. Behavioral trajectories
collected within such environments can be interpreted as realizations of
the perturbation--response mapping

\[u_{0:T - 1} \mapsto y_{0:T}.\]

Reconstruction then amounts to identifying a generative model capable of
reproducing this mapping across perturbation policies within the bounded
experimental domain.

Importantly, the role of the environment is not limited to delivering
stimuli. By systematically exploring the perturbation space, the
environment becomes an instrument for probing the internal dynamics of
behavior. Carefully designed perturbation policies can reveal nonlinear
responses, state-dependent transitions, and resilience limits that would
remain hidden under static experimental conditions. In this sense,
perturbation environments perform a function analogous to experimental
apparatus in the physical sciences: they provide the means through which
the underlying structure of the system becomes observable.

This perspective suggests a shift in the role of experimental design
within behavioral science. Rather than focusing solely on hypothesis
testing, experimental environments should be engineered to maximize the
informational value of perturbations for system identification. The
informativeness measure introduced in the previous section,

\[\Delta(\Pi_{\mathcal{D}}),\]

provides a formal criterion for evaluating the discriminatory power of
different perturbation families. Perturbation environments that generate
highly discriminative trajectories improve the feasibility and accuracy
of behavioral reconstruction.

Consequently, the design of perturbation environments becomes a central
methodological component of computational behavioral measurement. By
integrating programmable perturbations, multimodal sensing, and
reproducible experimental conditions, virtual environments provide the
infrastructure necessary to implement the reconstruction framework
described in the preceding section. They enable behavioral measurement
to move beyond static observation toward an active interrogation of the
dynamical mechanisms generating human behavior.

\section{Functional Equivalence and
Validation}\label{functional-equivalence-and-validation}

The reconstruction framework described in the previous sections raises a
fundamental methodological question: how can one determine whether a
reconstructed model faithfully captures the behavioral dynamics of the
system under investigation? Traditional model evaluation in behavioral
science typically relies on predictive accuracy or goodness-of-fit
metrics computed on the data used for model estimation. However, such
criteria are insufficient in the context of behavioral system
reconstruction. A model may reproduce the trajectories observed during
estimation without capturing the generative structure that produced
them. True reconstruction requires demonstrating that the model behaves
like the system itself when subjected to perturbations beyond those used
during training.

For this reason, validation within the present framework is defined in
terms of functional equivalence under novel perturbations. Rather than
evaluating whether a model fits observed data, we evaluate whether the
model generates trajectories that are statistically indistinguishable
from those produced by the behavioral system when exposed to previously
unseen perturbations.

Let \(\Pi_{\mathcal{D}}^{train}\) denote the family of perturbation
policies used to collect the training data for reconstruction, and let
\(\Pi_{\mathcal{D}}^{test}\) denote a disjoint set of perturbations
reserved for validation. The key requirement is that the perturbations
in the test set are not used during model estimation, thereby providing
a genuine out-of-distribution evaluation of the reconstructed model.

Given a perturbation sequence
\(u_{0:T - 1} \in \Pi_{\mathcal{D}}^{test}\) , the true behavioral
system generates a trajectory distribution

\[P_{\mathcal{B}}(y_{0:T} \mid u_{0:T - 1}),\]

while the reconstructed model \(\mathcal{M}_{\widehat{\theta}}\)
generates its own trajectory distribution

\[P_{\mathcal{M}_{\widehat{\theta}}}(y_{0:T} \mid u_{0:T - 1}).\]

The degree to which the reconstructed model captures the behavioral
dynamics of the system can therefore be evaluated by comparing these two
distributions. Let \(D( \cdot , \cdot )\) denote a discrepancy measure
defined over trajectory distributions (Kullback, S., \& Leibler, R. A.,
1951). Functional equivalence between the reconstructed model and the
behavioral system can then be expressed as the condition

\[\sup_{u \in \Pi_{\mathcal{D}}^{test}}D\left( P_{\mathcal{B}}( \cdot \mid u),P_{\mathcal{M}_{\widehat{\theta}}}( \cdot \mid u) \right) \leq \delta,\]

where \(\delta\) represents a tolerance threshold reflecting measurement
noise, model approximation error, and stochastic variability in
behavioral trajectories.

This definition captures the essential requirement that the
reconstructed model must reproduce the behavioral system's responses
across a family of perturbations rather than merely under the specific
conditions used for estimation. In practical terms, the model is
considered successful if an external observer exposed only to the
trajectory distributions generated by the model and by the real system
would be unable to reliably distinguish between them.

The choice of discrepancy measure \(D\) depends on the nature of the
behavioral trajectories and the stochastic structure of the system. In
deterministic or low-noise contexts, simple pointwise discrepancies
between trajectories may suffice. However, behavioral dynamics are
typically stochastic and heterogeneous across individuals and trials. In
such settings, comparison at the level of probability distributions over
trajectories is more appropriate (Goodfellow, I., Pouget-Abadie, J.,
Mirza, M., Xu, B., Warde-Farley, D., Ozair, S., ... \& Bengio, Y.,
2020). Distributional discrepancies such as the Wasserstein distance or
kernel-based divergences provide principled ways to compare stochastic
trajectory ensembles while accounting for variability in timing,
amplitude, and path structure.

An important implication of this validation framework is that
reconstruction becomes intrinsically tied to the notion of
out-of-distribution generalization. A model that perfectly reproduces
training trajectories may still fail when confronted with novel
perturbations. Such failures reveal that the reconstructed model
captures only superficial statistical regularities rather than the
underlying dynamical structure of behavior. Conversely, a model that
continues to generate indistinguishable trajectories under unseen
perturbations provides stronger evidence that the generative mechanisms
governing the behavioral system have been successfully captured.

Functional equivalence therefore shifts the focus of model validation
from retrospective fit to prospective behavioral indistinguishability.
This perspective is closely related to ideas from system identification
and generative modeling, where models are evaluated according to their
ability to reproduce the observable consequences of a system's internal
dynamics. In the present context, the criterion of indistinguishability
is evaluated over the space of perturbation--response trajectories
rather than static outputs.

This formulation also highlights the importance of the perturbation
environment in the validation process. The strength of the validation
test depends on the richness of the perturbation policies included in
\(\Pi_{\mathcal{D}}^{test}\) . If the test perturbations explore only a
narrow region of the perturbation space, models that are structurally
incorrect may still appear equivalent to the behavioral system.
Conversely, a diverse set of perturbations probing different regions of
the behavioral state space increases the likelihood that structural
discrepancies between the model and the system will become observable.

From this perspective, validation experiments can be interpreted as
stress tests of the reconstructed behavioral model. Perturbations that
push the system toward its dynamical boundaries---such as high cognitive
load, abrupt environmental changes, or social disruptions---are
particularly informative because they expose nonlinearities and state
transitions that may not be apparent under mild conditions. These
perturbations reveal whether the reconstructed model captures not only
the nominal behavior of the system but also its response to extreme or
rare conditions.

The validation framework also provides a natural way to characterize
reconstruction failures. Suppose that, for a given model class
\(\mathfrak{M}\) , the minimal achievable discrepancy between the
behavioral system and any model in the class is

\[\varepsilon^{\star}\mathfrak{(M,}\mathcal{D) =}\inf_{\theta \in \Theta}\sup_{u \in \Pi_{\mathcal{D}}^{test}}D\left( P_{\mathcal{B}}( \cdot \mid u),P_{\mathcal{M}_{\theta}}( \cdot \mid u) \right).\]

If \(\varepsilon^{\star}\) exceeds the tolerance level \(\delta\) , the
behavioral system cannot be reconstructed within the bounded
experimental domain using the specified model class. Rather than being
interpreted as a methodological failure, such outcomes reveal
fundamental limits on reconstructability. They indicate either that the
model class lacks sufficient expressive power to capture the dynamics of
the system or that the perturbation environment fails to generate
sufficiently informative trajectories.

In this way, validation becomes not merely a test of model performance
but a diagnostic tool for understanding the structure of behavioral
systems. Reconstruction success suggests that the system's dynamics are
adequately captured within the chosen model class and perturbation
domain. Reconstruction failure, on the other hand, points toward deeper
complexity in the behavioral process or toward insufficient perturbation
richness in the experimental environment.

Ultimately, the concept of functional equivalence establishes a rigorous
criterion for behavioral system reconstruction. By requiring
indistinguishability under novel perturbations, it ensures that
reconstructed models capture the generative mechanisms underlying
behavioral dynamics rather than simply fitting observed data. This
principle provides the methodological foundation for a
perturbation-based approach to behavioral measurement in which models
are evaluated according to their ability to reproduce the observable
consequences of human behavior across a controlled space of
environmental interventions.

\section{Perturbation Informativeness and Experimental
Design}\label{perturbation-informativeness-and-experimental-design}

The reconstruction framework presented in the preceding sections
emphasizes that behavioral dynamics can only be identified through
sufficiently informative perturbations. While functional equivalence
provides a criterion for evaluating reconstructed models, successful
reconstruction depends critically on the design of perturbation policies
capable of revealing the internal structure of the behavioral system.
Experimental design therefore becomes a central methodological component
of behavioral reconstruction.

In classical experimental psychology, experimental manipulations are
typically selected to test specific hypotheses about cognitive or
affective processes. Conditions are designed to isolate particular
effects, often through factorial manipulations or controlled contrasts
between stimulus categories. Although this strategy has been highly
effective for hypothesis testing, it is not optimized for system
identification. From the perspective of reconstruction, the objective is
not simply to detect differences between conditions but to generate
behavioral trajectories that expose the dynamical structure of the
system (Strogatz, S. H., 2024).

Within the present framework, perturbations function as identification
inputs analogous to those used in control theory and system
identification. Let \(\Pi_{\mathcal{D}}\) denote the family of
admissible perturbation policies defined within the bounded experimental
domain. Each policy \(\pi \in \Pi_{\mathcal{D}}\) generates a sequence
of perturbations \(u_{0:T - 1}\) that interacts with the behavioral
system and produces a corresponding trajectory distribution

\[P_{\mathcal{B}}(y_{0:T} \mid u_{0:T - 1}).\]

Different perturbation policies may elicit trajectories that are more or
less informative about the underlying system dynamics (Strogatz, S. H.,
2024). If two distinct candidate models generate indistinguishable
trajectories under a given perturbation policy, that policy provides no
information for discriminating between them. Conversely, perturbations
that produce strongly divergent trajectories across models reveal
structural differences in their underlying dynamics.

To formalize this intuition, consider two candidate models
\(\mathcal{M}_{\theta_{1}}\) and \(\mathcal{M}_{\theta_{2}}\) belonging
to the model class \(\mathfrak{M}\) . The ability of a perturbation
sequence to discriminate between these models can be measured by the
discrepancy between their induced trajectory distributions:

\[D\left( P_{\mathcal{M}_{\theta_{1}}}( \cdot \mid u),P_{\mathcal{M}_{\theta_{2}}}( \cdot \mid u) \right).\]

A perturbation family \(\Pi_{\mathcal{D}}\) is informative if, for any
pair of distinct parameter values, there exists at least one
perturbation policy capable of producing distinguishable behavioral
trajectories. This property can be captured by defining the
discriminatory power of the perturbation family as

\[\Delta(\Pi_{\mathcal{D}}) = \inf_{\theta_{1} \neq \theta_{2}}\sup_{u \in \Pi_{\mathcal{D}}}D\left( P_{\mathcal{M}_{\theta_{1}}}( \cdot \mid u),P_{\mathcal{M}_{\theta_{2}}}( \cdot \mid u) \right).\]

If \(\Delta(\Pi_{\mathcal{D}}) > 0\) , distinct models generate
distinguishable trajectory distributions under at least one admissible
perturbation. In such cases, the perturbation family contains sufficient
information to enable reconstruction in principle. Conversely, if
\(\Delta(\Pi_{\mathcal{D}}) = 0\) , there exist distinct models that
remain indistinguishable under all admissible perturbations, making
reconstruction impossible within the given experimental domain.

This formulation reveals that the feasibility of behavioral
reconstruction depends not only on the choice of model class but also on
the design of the perturbation environment. Even highly expressive
models cannot be reliably identified if the perturbation policies fail
to probe the relevant dimensions of the behavioral state space.

From this perspective, the design of perturbation environments becomes
an optimization problem. Let \(\mathcal{A}\) denote the set of feasible
perturbation policies that can be implemented within the experimental
infrastructure. The objective is to identify the perturbation family
that maximizes discriminatory power across the model class. Formally,
the optimal perturbation family can be defined as

\[\Pi_{\mathcal{D}}^{\star} = arg\max_{\Pi \in \mathcal{A}}\Delta(\Pi).\]

Perturbation policies selected according to this criterion generate
behavioral trajectories that are maximally informative for
distinguishing between competing models of behavioral dynamics.

In practice, optimal perturbation design may involve balancing several
competing objectives. Highly informative perturbations may also induce
extreme behavioral responses that compromise ecological validity or
participant comfort. Furthermore, certain perturbations may be
infeasible due to experimental constraints or ethical considerations. As
a result, optimal perturbation design typically involves maximizing
informational value subject to feasibility and safety constraints.

An important advantage of programmable perturbation environments is
their ability to implement adaptive perturbation policies. Rather than
predefining a fixed sequence of perturbations, the environment can
dynamically adjust perturbations in response to the evolving behavior of
the participant. Let \(s_{t}\) denote the environmental state and
\(y_{t}\) the observed behavioral output at time \(t\) . An adaptive
perturbation policy can then be expressed as

\[u_{t} = \pi(s_{t},y_{0:t}),\]

allowing perturbations to depend on the history of observed behavior.
Such policies can actively explore regions of the behavioral state space
that are most informative for system identification. For example, if the
current behavioral trajectory suggests that the system is near a
stability boundary, the perturbation policy may introduce additional
stressors to reveal nonlinear transitions or resilience limits.

Adaptive perturbation design transforms the experimental environment
into an active interrogation mechanism for behavioral systems. Instead
of passively presenting stimuli, the environment interacts with the
participant in a closed loop, continuously adjusting perturbations to
maximize the informational value of the resulting trajectories.

This approach aligns behavioral experimentation with principles from
optimal experimental design and active learning. By systematically
selecting perturbations that maximize information gain about model
parameters or system structure, the experimental process itself becomes
part of the reconstruction algorithm. Data collection and model
identification are therefore integrated within a unified framework.

The implications of this perspective extend beyond the design of
individual experiments. Perturbation environments can be progressively
refined as reconstruction progresses. Early experiments may employ broad
exploratory perturbations to identify coarse features of the behavioral
system. Subsequent experiments can then focus on perturbations that
discriminate between competing models or refine estimates of critical
dynamical parameters. In this way, the experimental design evolves
iteratively alongside the reconstruction process.

Ultimately, perturbation informativeness provides a formal bridge
between experimental methodology and computational modeling in
behavioral science. By treating perturbations as identification inputs
and by designing experimental environments that maximize their
informational value, behavioral research can move beyond static
observation toward an active reconstruction of the dynamical mechanisms
that generate human behavior.

\section{Identifiability and Limits of Behavioral
Reconstruction}\label{identifiability-and-limits-of-behavioral-reconstruction}

The reconstruction framework described in the preceding sections
establishes a formal methodology for identifying behavioral systems from
perturbation--response trajectories. However, the feasibility of
reconstruction depends on fundamental theoretical conditions. Even in
the presence of rich experimental data, it is possible that distinct
models generate indistinguishable observable trajectories. In such cases
the behavioral system cannot be uniquely reconstructed. Understanding
the conditions under which reconstruction is possible therefore requires
a careful analysis of identifiability and reconstructability limits.

Identifiability concerns whether the parameters or structures of a model
can be uniquely inferred from observable data. Within the present
framework, the observable data consist of trajectory distributions
generated under admissible perturbation policies. Let
\(\mathfrak{M = \{}\mathcal{M}_{\theta}:\theta \in \Theta\}\) denote the
candidate model class. Each model induces a family of trajectory
distributions

\[P_{\mathcal{M}_{\theta}}(y_{0:T} \mid u_{0:T - 1})\]

for perturbation sequences \(u_{0:T - 1}\) drawn from the admissible
perturbation family \(\Pi_{\mathcal{D}}\) . Reconstruction is possible
only if distinct parameter values generate distinguishable trajectory
distributions across the perturbation space.

This condition can be formalized through the concept of structural
identifiability. The model class \(\mathfrak{M}\) is structurally
identifiable on the bounded experimental domain \(\mathcal{D}\) if

\[P_{\mathcal{M}_{\theta_{1}}}( \cdot \mid u) = P_{\mathcal{M}_{\theta_{2}}}( \cdot \mid u)\quad\forall u \in \Pi_{\mathcal{D}}\quad \Rightarrow \quad\theta_{1} = \theta_{2}.\]

This condition states that two parameter values producing identical
trajectory distributions under all admissible perturbations must in fact
be identical. If structural identifiability fails, multiple models
remain observationally indistinguishable regardless of the amount of
available data.

Structural identifiability is a property of the model class and
perturbation environment rather than the dataset itself. Even with
infinite data, reconstruction is impossible if the perturbation policies
fail to reveal differences between candidate models. The informativeness
criterion introduced earlier provides a complementary perspective on
this issue. A perturbation family capable of separating all models
within the candidate class ensures that the identifiability condition
can, in principle, be satisfied.

In practice, behavioral data are finite and subject to measurement
noise, individual variability, and stochastic fluctuations in behavioral
trajectories. These factors give rise to the concept of practical
identifiability. Even when a model class is structurally identifiable,
parameter estimation may remain unstable if the available data do not
provide sufficient information to constrain the parameter space.

Let \({\widehat{\theta}}_{N}\) denote an estimator obtained from \(N\)
perturbation--response trajectories. Practical identifiability requires
that the uncertainty associated with the estimator decreases as the
amount of data increases. Formally, the estimator should satisfy a
consistency condition such as

\[{\widehat{\theta}}_{N} \rightarrow \theta^{\star}\quad\text{as}\quad N \rightarrow \infty,\]

where \(\theta^{\star}\) denotes the parameter value corresponding to
the true behavioral system within the model class. When this condition
holds, increasing the number or diversity of perturbation trajectories
improves the accuracy of reconstruction.

However, even under ideal estimation conditions, reconstruction remains
constrained by the expressive power of the model class. The true
behavioral system may lie outside the candidate model family, in which
case perfect reconstruction is impossible. The discrepancy between the
true system and the best approximating model defines the intrinsic
reconstruction error

\[\varepsilon^{\star}\mathfrak{(M,}\mathcal{D) =}\inf_{\theta \in \Theta}\sup_{u \in \Pi_{\mathcal{D}}}D\left( P_{\mathcal{B}}( \cdot \mid u),P_{\mathcal{M}_{\theta}}( \cdot \mid u) \right).\]

This quantity represents the smallest achievable discrepancy between the
behavioral system and any model within the candidate class over the
admissible perturbation domain. If \(\varepsilon^{\star}\) exceeds the
tolerance level required for functional equivalence, the behavioral
system cannot be reconstructed within the specified model class.

The concept of intrinsic reconstruction error highlights a fundamental
trade-off in behavioral modeling. Model classes with limited complexity
may fail to capture essential dynamical features of the system, leading
to large reconstruction errors. Conversely, highly expressive model
classes may introduce identifiability challenges, as many parameter
configurations may generate similar observable trajectories. Effective
reconstruction therefore requires balancing expressive power with
identifiability constraints.

An additional limitation arises from the bounded nature of experimental
domains. Behavioral systems operate across a vast range of environmental
conditions, many of which cannot be reproduced experimentally.
Reconstruction claims must therefore be interpreted relative to the
perturbation domain \(\mathcal{D}\) . A model that achieves functional
equivalence within the experimental domain may still fail when exposed
to perturbations outside that domain.

This observation motivates the notion of domain-specific
reconstructability. Let \(\mathcal{D}_{1}\) and \(\mathcal{D}_{2}\)
denote two distinct perturbation domains. A behavioral system may be
reconstructable within one domain but not the other if the perturbations
available in the second domain fail to reveal critical aspects of the
system dynamics (Strogatz, S. H., 2024). Reconstruction is therefore not
an absolute property of a behavioral system but a relational property
defined with respect to the experimental environment and the chosen
model class.

These considerations lead to an important conceptual conclusion:
reconstruction failures are themselves scientifically informative. If no
model within a candidate class achieves functional equivalence under the
admissible perturbations, this indicates that either the model class
lacks sufficient expressive power or that the perturbation environment
does not sufficiently probe the behavioral dynamics. In both cases, the
failure provides guidance for refining the modeling framework or
redesigning the experimental environment.

The study of identifiability and reconstructability limits therefore
plays a central role in the proposed measurement paradigm. Rather than
assuming that behavioral systems can always be faithfully modeled, the
framework explicitly characterizes the conditions under which
reconstruction is feasible and the circumstances under which it fails.
This perspective transforms reconstruction from a purely methodological
task into a scientific inquiry into the structural properties of
behavioral systems and the experimental environments used to interrogate
them.

\section{Special Case: Linear Gaussian Behavioral
Systems}\label{special-case-linear-gaussian-behavioral-systems}

The reconstruction framework developed in the previous sections was
formulated at a high level of generality, treating behavioral systems as
stochastic controlled dynamical systems whose trajectory distributions
must be inferred from perturbation--response data. While this general
formulation allows for a wide range of nonlinear and stochastic models,
it is useful to consider specific model classes in which reconstruction
can be analyzed more concretely. Linear Gaussian state-space systems
provide a particularly instructive example because they combine
analytical tractability with sufficient expressive power to capture a
variety of behavioral dynamics (Kitagawa, G., \& Gersch, W., 1996).

In a linear Gaussian behavioral system, the latent behavioral state
evolves according to a linear dynamical equation driven by external
perturbations and stochastic noise. Let \(x_{t} \in \mathbb{R}^{n}\)
denote the latent state vector and \(u_{t} \in \mathbb{R}^{m}\) the
perturbation applied at time \(t\) . The system dynamics are described
by

\[x_{t + 1} = Ax_{t} + Bu_{t} + w_{t},\]

where \(A\) is an \(n \times n\) state transition matrix, \(B\) is an
\(n \times m\) perturbation coupling matrix, and \(w_{t}\) is a Gaussian
process noise term with distribution

\[w_{t}\mathcal{\sim N(}0,Q).\]

The observable behavioral output \(y_{t} \in \mathbb{R}^{p}\) is
generated from the latent state through the observation equation

\[y_{t} = Cx_{t} + v_{t},\]

where \(C\) is a \(p \times n\) observation matrix and \(v_{t}\)
represents measurement noise with distribution

\[v_{t}\mathcal{\sim N(}0,R).\]

Together, these equations define a stochastic linear dynamical system
characterized by the parameter set

\[\theta = (A,B,C,Q,R).\]

Within this model class, perturbations influence behavioral trajectories
through the matrix \(B\) , which determines how environmental
interventions affect the evolution of the latent behavioral state. The
matrices \(A\) and \(Q\) describe the intrinsic dynamics of the
behavioral system, including stability properties and stochastic
fluctuations, while \(C\) and \(R\) determine how latent states are
reflected in observable behavioral signals.

The reconstruction problem in this setting reduces to estimating the
parameter set \(\theta\) from perturbation--response trajectories. Given
a dataset of observed trajectories and perturbation sequences

\[\mathcal{D}_{N} = \left\{ (u_{0:T_{i} - 1}^{(i)},y_{0:T_{i}}^{(i)}) \right\}_{i = 1}^{N},\]

parameter estimation can be performed through likelihood maximization or
Bayesian inference. Because the system is linear and Gaussian, the
likelihood of each trajectory can be computed efficiently using Kalman
filtering and smoothing algorithms. These methods provide estimates of
the latent state sequence as well as the parameters governing the system
dynamics.

Within the linear Gaussian framework, the distribution of behavioral
trajectories can be derived analytically. For a given perturbation
sequence \(u_{0:T - 1}\) , the trajectory \(y_{0:T}\) follows a
multivariate Gaussian distribution whose mean and covariance are
determined by the system matrices and noise covariances. This analytical
structure allows discrepancies between the behavioral system and
reconstructed models to be evaluated using closed-form expressions for
trajectory distributions.Perturbations play a particularly transparent
role in this model class. Because the latent dynamics evolve linearly
with respect to \(u_{t}\) , different perturbation sequences produce
predictable changes in the mean trajectory of the behavioral system. If
the perturbation matrix \(B\) has full rank and the perturbation
sequences explore sufficiently diverse directions in the perturbation
space, the resulting trajectories provide strong constraints on the
system parameters. In contrast, if perturbations remain confined to a
low-dimensional subspace, certain parameters may become unidentifiable
because different parameter configurations can produce identical
observable trajectories (Ljung, 2010).

These considerations illustrate how perturbation design interacts with
identifiability even in relatively simple dynamical systems. For
example, consider the controllability matrix

\[\mathcal{C =}\begin{bmatrix}
B & AB & A^{2}B & \ldots & A^{n - 1}B
\end{bmatrix}.\]

If this matrix has full rank, the perturbation inputs can influence all
dimensions of the latent state space. In such cases, perturbations
provide a powerful mechanism for probing the internal dynamics of the
system. Conversely, if the controllability matrix is rank-deficient,
certain latent dimensions remain inaccessible to perturbation, limiting
the ability to reconstruct the corresponding behavioral dynamics.

The linear Gaussian model also allows a clear illustration of the
concept of functional equivalence introduced earlier. Suppose a
reconstructed model \(\mathcal{M}_{\widehat{\theta}}\) has parameter
estimates
\(\widehat{A},\widehat{B},\widehat{C},\widehat{Q},\widehat{R}\) . For
any perturbation sequence \(u_{0:T - 1}\) , the model generates a
predicted trajectory distribution

\[P_{\mathcal{M}_{\widehat{\theta}}}(y_{0:T} \mid u_{0:T - 1}\mathcal{) = N(}\mu_{\widehat{\theta}},\Sigma_{\widehat{\theta}}),\]

where \(\mu_{\widehat{\theta}}\) and \(\Sigma_{\widehat{\theta}}\)
denote the mean and covariance implied by the estimated parameters.
Functional equivalence can then be evaluated by comparing these
predicted distributions to those generated by the behavioral system
under novel perturbations.

Because Gaussian distributions admit closed-form divergence measures,
discrepancies between trajectory distributions can be computed
analytically. For example, the Wasserstein distance between two Gaussian
trajectory distributions depends only on the differences between their
means and covariance matrices (Peyré, G., \& Cuturi, M., 2019). This
property enables efficient evaluation of functional equivalence across
large sets of perturbation trajectories.

Although linear Gaussian systems represent a simplified model class,
they provide an instructive bridge between the abstract reconstruction
framework and practical behavioral modeling. Many existing approaches to
modeling cognitive and affective dynamics can be expressed as special
cases of linear state-space systems or as nonlinear extensions thereof
(Waugh, C. E., \& Kuppens, P., 2021; Valenza, G., Lanata, A., \&
Scilingo, E. P., 2011; Markov, K., Matsui, T., Septier, F., \& Peters,
G., 2015; Hollenstein, T., \& Lewis, M. D., 2006; Hamaker, E. L.,
Ceulemans, E., Grasman, R. P., \& Tuerlinckx, F., 2015). Moreover, the
analytical tools developed for linear dynamical systems---such as
controllability analysis, Kalman filtering, and optimal experiment
design---provide a foundation for implementing perturbation-based
behavioral reconstruction in real experimental settings.

More broadly, the linear Gaussian case illustrates how the general
reconstruction framework can be instantiated within concrete model
classes. By analyzing identifiability, perturbation informativeness, and
functional equivalence within specific dynamical models, researchers can
develop practical reconstruction strategies tailored to particular
behavioral domains.

The considerations above suggest that reconstruction critically depends
on the informativeness of perturbations. This dependence can be made
explicit in a simple controlled setting, where identifiability
conditions can be directly observed.

\section{Simulation}\label{simulation}

To demonstrate the practical feasibility of the proposed framework, we
consider a simple linear Gaussian dynamical system with known
parameters, where behavioral dynamics are generated according to a
controlled state-space model and perturbations act as external inputs.

A dataset of perturbation--response trajectories is generated under two
distinct conditions: (i) a rich perturbation regime with sufficient
variability, and (ii) a low-variability regime with nearly constant
inputs. In both cases, models are estimated from the training data using
standard system identification techniques and subsequently evaluated
under a set of unseen test perturbations.

Functional equivalence is assessed by comparing empirical trajectory
distributions obtained from the true system and from the reconstructed
model using a Wasserstein discrepancy measure. Results show a clear
divergence between the two regimes (Figure 2). Under rich perturbations,
the reconstructed model produces trajectories that closely match the
true system dynamics, resulting in low discrepancy and statistical
indistinguishability across test conditions. In contrast, under
low-variability perturbations, the reconstruction fails to capture the
system behavior, leading to substantially higher discrepancy and
degenerate trajectory predictions.

These findings demonstrate that the success of behavioral system
reconstruction critically depends on the informativeness of the
perturbations. In the absence of sufficiently rich inputs, the
estimation problem becomes not only inaccurate but effectively
ill-posed, as the available data do not contain enough information to
identify the underlying system.

\begin{figure}[htbp]
\centering
\includegraphics[width=\textwidth]{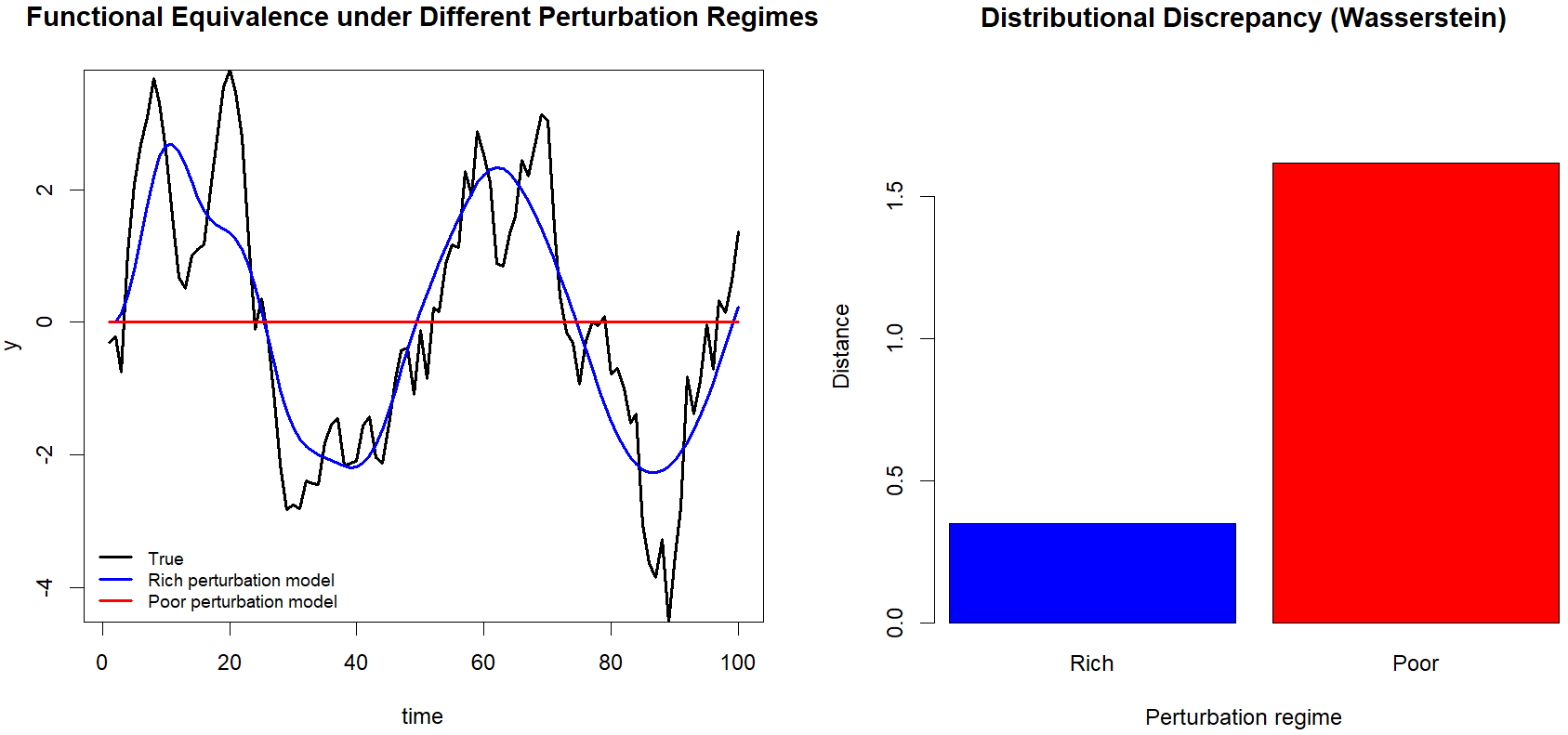}
\caption{Functional equivalence under different perturbation regimes. Left: Example trajectories generated by the true system (black), the model reconstructed under rich perturbations (blue), and the model reconstructed under poor perturbations (red), evaluated under an unseen test input. Right: Distributional discrepancy (Wasserstein distance) between true and model-generated trajectories. Rich perturbations yield low discrepancy, while poor perturbations lead to substantial deviations, highlighting the role of perturbation informativeness in behavioral system reconstruction. Numerical values: Wasserstein distance $= 0.35$ (rich) vs $1.62$ (poor).}
\end{figure}

\section{Behavioral Domains and Empirical
Instantiation}\label{behavioral-domains-and-empirical-instantiation}

Human behavioral processes manifest across diverse domains characterized
by different dynamical structures, temporal scales, and forms of
environmental interaction. Emotional regulation, cognitive flexibility,
stress reactivity, and social coordination all involve behavioral states
that evolve over time and respond to external perturbations. Within the
reconstruction framework introduced in this paper, these domains can be
conceptualized as stochastic dynamical systems whose internal states are
probed through controlled perturbations and whose dynamics are inferred
from the resulting behavioral trajectories. Crucially, perturbations are
not introduced merely as experimental stimuli but as interventions
designed to induce transitions between latent behavioral states, thereby
revealing the structure of the underlying system dynamics (Strogatz, S.
H., 2024). Different behavioral domains therefore require perturbation
environments capable of eliciting domain-relevant state transitions
while producing trajectories that can be modeled using suitable
dynamical representations. Table 1 provides illustrative examples
linking behavioral domains, targeted state transitions, perturbation
environments, and candidate model classes that may be used to
reconstruct the corresponding behavioral dynamics (Kloeden, P. E., \&
Pearson, R. A., 1977; Platen, E., \& Bruti-Liberati, N., 2010;
Ghahramani, Z., \& Hinton, G. E., 1996; Eddy, S. R., 2004; Wiggins, S.,
2003; Ghaffarzadegan, N., Majumdar, A., Williams, R., \& Hosseinichimeh,
N., 2024; Durbin, J., \& Koopman, S. J., 2012).

\begin{center}\textbf{Table 1.} Illustrative instantiations of perturbation-based behavioral reconstruction.\end{center}

\begin{longtable}[]{@{}
  >{\raggedright\arraybackslash}p{(\columnwidth - 10\tabcolsep) * \real{0.1667}}
  >{\raggedright\arraybackslash}p{(\columnwidth - 10\tabcolsep) * \real{0.1667}}
  >{\raggedright\arraybackslash}p{(\columnwidth - 10\tabcolsep) * \real{0.1667}}
  >{\raggedright\arraybackslash}p{(\columnwidth - 10\tabcolsep) * \real{0.1666}}
  >{\raggedright\arraybackslash}p{(\columnwidth - 10\tabcolsep) * \real{0.1666}}
  >{\raggedright\arraybackslash}p{(\columnwidth - 10\tabcolsep) * \real{0.1666}}@{}}
\toprule\noalign{}
\begin{minipage}[b]{\linewidth}\raggedright
\textbf{Behavioral domain}
\end{minipage} & \begin{minipage}[b]{\linewidth}\raggedright
\textbf{Targeted state transitions}
\end{minipage} & \begin{minipage}[b]{\linewidth}\raggedright
\textbf{Perturbation environment}
\end{minipage} & \begin{minipage}[b]{\linewidth}\raggedright
\textbf{Example VR manipulation}
\end{minipage} & \begin{minipage}[b]{\linewidth}\raggedright
\textbf{Observable behavioral trajectories}
\end{minipage} & \begin{minipage}[b]{\linewidth}\raggedright
\textbf{Illustrative model classes}
\end{minipage} \\
\midrule\noalign{}
\endhead
\bottomrule\noalign{}
\endlastfoot
Affective dynamics & Transitions between emotional regimes (e.g.,
relaxation $\rightarrow$ stress $\rightarrow$ recovery) & Emotional perturbation protocol & VR
scenarios manipulating social evaluation, environmental threat, or
calming environments & Continuous affect ratings, heart-rate
variability, gaze dynamics, body movement & Stochastic differential
equations; nonlinear dynamical systems \\
Cognitive flexibility & Transitions between task strategies or cognitive
control states & Adaptive task environment & VR tasks with rule
switches, changing contingencies, or unexpected feedback & Decision
sequences, response times, error dynamics & Hidden Markov models;
switching state-space models \\
Stress reactivity and resilience & Transitions between stable
performance and breakdown under stress & Escalating challenge
environment & Increasing time pressure, environmental noise, or
unpredictable events & Performance degradation trajectories,
physiological stress indicators, recovery dynamics & Nonlinear dynamical
systems; stochastic stability models \\
Social interaction dynamics & Transitions between coordination patterns
among interacting agents & Multi-agent interaction environment &
Cooperative or competitive VR scenarios with adaptive agent behavior &
Joint behavioral trajectories, communication timing, coordination
metrics & Agent-based models; coupled dynamical systems \\
\end{longtable}

A first class of behavioral systems concerns affective dynamics.
Emotional states evolve continuously over time and are influenced by
both external events and internal regulatory processes. Let \(x_{t}\)
represent a low-dimensional latent state describing the affective
configuration of an individual at time \(t\) . Environmental events,
social interactions, or cognitive demands may act as perturbations
\(u_{t}\) that modify this state. Observable outputs \(y_{t}\) may
include self-reported affect ratings, physiological measures such as
heart rate variability, or behavioral indicators such as facial
expression and movement patterns.

Within this domain, reconstruction seeks to identify dynamical models
capable of capturing how affective states evolve under perturbations and
how individuals recover from disturbances. Perturbation environments may
involve controlled emotional stimuli, social feedback manipulations, or
task demands designed to induce fluctuations in affective states (Ljung,
2010). The resulting trajectory data allow researchers to estimate
dynamical properties such as stability, inertia, and resilience of
affective systems.

A second important domain concerns cognitive flexibility and adaptive
behavior. Cognitive systems must constantly adapt to changing
environmental demands, switching strategies when task contingencies
change. In such contexts, perturbations may correspond to shifts in task
rules, introduction of conflicting stimuli, or modifications of reward
structures. The latent behavioral state may represent internal
representations of task structure, cognitive control allocation, or
belief states about environmental contingencies.

Reconstruction in this domain aims to identify models capable of
reproducing how individuals adapt to perturbations in task structure.
Behavioral trajectories may include sequences of decisions, response
times, and action patterns that evolve as individuals learn or adapt to
new conditions. By systematically manipulating environmental
contingencies, perturbation environments can reveal how cognitive
systems transition between stable strategies and how they recover from
disruptive changes.

A third domain involves stress reactivity and resilience. Human
behavioral systems exhibit complex responses to environmental stressors,
including physiological, cognitive, and affective components.
Perturbations in this context may involve time pressure, uncertainty,
social evaluation, or unexpected environmental disruptions. Observable
outputs may include behavioral performance, physiological stress
indicators, and patterns of decision making.

Within the reconstruction framework, such perturbations allow
researchers to probe the stability boundaries of behavioral systems.
Mild perturbations may produce only transient deviations from baseline
behavior, while stronger perturbations may push the system toward
nonlinear transitions or breakdowns in adaptive functioning. By
analyzing how behavioral trajectories evolve under increasing
perturbation intensity, reconstruction methods can estimate resilience
properties of the system and identify critical transition points.

Social interaction dynamics provide another domain in which the
reconstruction paradigm becomes particularly relevant. Social behavior
often involves coupled dynamical systems in which multiple agents
interact and influence one another. The behavioral state of each
participant evolves not only in response to environmental perturbations
but also in response to the behavior of other agents. Perturbations in
this context may involve modifications of interaction rules,
introduction of cooperative or competitive incentives, or changes in
communication channels.

In such systems, reconstruction must account for the mutual influence
between interacting agents. Behavioral trajectories become joint
trajectories describing the evolution of multiple behavioral states over
time. Identifying the underlying dynamical structure requires
perturbation environments capable of manipulating interaction structures
while recording the resulting behavioral responses of all participants.

Across these domains, the reconstruction framework provides a unifying
perspective on behavioral measurement. Rather than treating emotional
states, cognitive strategies, or social behaviors as isolated variables,
the framework treats them as dynamical systems whose properties can be
inferred from perturbation--response trajectories. Different domains may
require different model classes and perturbation environments, but the
underlying methodological principle remains the same: behavioral systems
are reconstructed by systematically probing their responses to
controlled interventions.

Virtual environments and programmable experimental platforms play a
crucial role in enabling such empirical instantiations. These
environments allow perturbations to be implemented along multiple
dimensions simultaneously, including sensory, cognitive, and social
variables. At the same time, they provide high-resolution measurements
of behavioral trajectories through multimodal sensing technologies. The
integration of controlled perturbations with continuous behavioral
measurement creates the conditions necessary for applying the
reconstruction methodology across diverse behavioral domains.

Importantly, the framework does not assume that all behavioral systems
are equally reconstructable. Different domains may exhibit varying
degrees of dynamical complexity and stochastic variability. Some
behavioral processes may be well approximated by relatively simple
dynamical models, while others may require highly nonlinear or
high-dimensional representations. By applying the reconstruction
framework across multiple behavioral domains, researchers can
empirically investigate the limits of reconstructability and identify
which aspects of human behavior can be faithfully modeled within bounded
experimental environments.

In this way, the empirical instantiation of the reconstruction framework
serves not only to model specific behavioral systems but also to advance
a broader scientific objective: understanding the extent to which human
behavior can be represented as a generative dynamical process whose
observable trajectories can be reproduced under controlled
perturbations.

\section{Relation to Psychometrics and Computational
Modeling}\label{relation-to-psychometrics-and-computational-modeling}

The reconstruction framework proposed in this work occupies an
intermediate position between two established traditions in behavioral
science: classical psychometrics and contemporary computational modeling
(Marsman, M., Waldorp, L., \& Borsboom, D., 2023). While both traditions
aim to formalize and explain behavioral data, they differ substantially
in their assumptions about what constitutes a measurement and how
behavioral systems should be represented. The present framework
integrates elements from both perspectives while extending them toward a
dynamical and intervention-based conception of behavioral measurement.

Classical psychometrics has historically focused on the measurement of
latent traits inferred from patterns of observed responses. In this
paradigm, behavioral observations are typically treated as
manifestations of stable underlying attributes such as abilities,
attitudes, or personality traits. Measurement models, such as factor
analysis or item response theory, establish probabilistic relationships
between latent variables and observed responses, allowing the estimation
of latent scores from observed data. These approaches have proven
extremely powerful for constructing reliable instruments and for
analyzing large-scale behavioral datasets. However, they generally
assume that the underlying latent structure is static or slowly varying,
and they rely on repeated observations of similar stimuli rather than on
controlled perturbations of the behavioral system.

The reconstruction framework introduced in this paper adopts a different
perspective on behavioral measurement. Instead of treating latent
variables as static attributes inferred from aggregated responses, it
treats behavior as the observable output of an evolving dynamical
system. Within this view, measurement consists not merely of estimating
latent scores but of identifying the generative mechanisms that produce
behavioral trajectories under controlled perturbations. Observable data
are therefore interpreted as realizations of a perturbation-response
mapping that reflects the internal dynamics of the behavioral system.
The object of measurement becomes the trajectory law induced by the
system rather than a static parameter associated with an individual.

This shift from latent trait estimation to dynamical reconstruction does
not invalidate the contributions of classical psychometrics. On the
contrary, many psychometric models can be interpreted as special cases
of the broader dynamical framework considered here. For example, item
response models may be viewed as simplified observation models in which
latent variables remain constant during the measurement process and
perturbations correspond to the presentation of test items. In this
sense, psychometric measurement can be understood as a limiting case of
behavioral reconstruction in which the temporal dimension and dynamical
interactions are suppressed. Extending psychometric models to
incorporate temporal dynamics, state transitions, and controlled
perturbations therefore represents a natural generalization of existing
measurement theory.

At the same time, the proposed framework connects naturally with recent
developments in computational modeling of behavior. Over the past two
decades, computational approaches have increasingly represented
cognitive and behavioral processes using dynamical systems,
reinforcement learning models, Bayesian inference mechanisms, and
agent-based simulations. These models aim to capture the generative
processes underlying behavior rather than merely describing statistical
associations among observed variables. Computational modeling therefore
shares with the reconstruction framework the goal of explaining behavior
in terms of mechanistic processes.

However, computational models are often developed primarily as
theoretical constructs or explanatory tools rather than as measurement
instruments. Model evaluation typically relies on predictive accuracy or
goodness-of-fit metrics applied to observational data. While these
criteria are useful for comparing models, they do not necessarily
establish that a model faithfully reproduces the behavioral dynamics of
the system under novel conditions. The reconstruction framework
addresses this limitation by introducing the criterion of functional
equivalence under controlled perturbations. A reconstructed model is
considered valid only if it produces trajectory distributions
indistinguishable from those generated by the behavioral system when
exposed to previously unseen perturbations.

This emphasis on perturbation-based validation aligns the framework with
principles from system identification and experimental control theory.
In these disciplines, models are not evaluated solely on their ability
to fit observed data but on their ability to reproduce the system's
response to external inputs across a range of conditions. By adopting a
similar perspective, behavioral reconstruction transforms experimental
environments into identification instruments capable of probing the
internal dynamics of human behavior.

More fundamentally, in the absence of sufficiently informative
perturbations, the reconstruction problem becomes not only less accurate
but effectively ill-posed, as the available data do not contain enough
information to identify the underlying system dynamics.

The resulting perspective suggests a synthesis of psychometric
measurement and computational modeling. Psychometrics contributes
rigorous statistical frameworks for inference and uncertainty
quantification, while computational modeling provides mechanistic
representations of behavioral processes. The perturbation-based
reconstruction framework integrates these traditions within a unified
methodological structure in which behavioral systems are interrogated
through controlled interventions and reconstructed as generative
dynamical models. This integration opens the possibility of developing
measurement instruments that do not merely estimate latent traits but
actively reconstruct the dynamical mechanisms governing human behavior.

\section{Toward a Metrology of Human
Behavior}\label{toward-a-metrology-of-human-behavior}

The framework developed in this paper suggests a broader conceptual
shift in how behavioral measurement can be understood. Traditionally,
the measurement of psychological constructs has relied on statistical
associations between observed responses and latent variables. Although
these approaches have produced powerful tools for assessing cognitive
abilities, personality traits, and attitudes, they rarely establish
measurement in the strict sense used in the physical sciences. In most
cases, behavioral constructs are inferred from patterns of responses
rather than directly tied to experimentally controlled transformations
of the system under study. The perturbation-based reconstruction
framework opens the possibility of moving toward a more rigorous
conception of behavioral measurement that parallels the logic of
metrology.

In the physical sciences, metrology refers to the science of measurement
and involves the systematic definition of quantities, units, and
measurement procedures that yield reproducible and comparable results
across laboratories and contexts. Measurements are typically grounded in
controlled interactions between a measuring instrument and the physical
system of interest. Through these interactions, observable responses
reveal properties of the underlying system, allowing researchers to
estimate quantities that characterize its structure or dynamics. The
validity of a measurement procedure is established by demonstrating that
it reliably captures invariant properties of the system across repeated
observations and varying experimental conditions.

A similar perspective can be adopted for the study of human behavior.
Within the reconstruction framework, experimental environments (such as
adaptive tasks, immersive virtual reality scenarios, or interactive
social simulations), function as measurement instruments designed to
probe the dynamics of behavioral systems. Controlled perturbations serve
the role of measurement operations, systematically interacting with the
behavioral system and eliciting responses that reveal its internal
structure. Observable trajectories, such as sequences of actions,
physiological responses, or interaction patterns, correspond to the
measurement signals generated by these operations.

From this viewpoint, behavioral measurement no longer consists solely of
assigning numerical scores to individuals but of identifying the
dynamical mechanisms that govern behavioral responses to perturbations.
The quantities of interest are therefore not limited to static latent
variables but may include dynamical parameters, transition structures,
stability properties, or other characteristics of the underlying
behavioral system. In this sense, behavioral reconstruction aims to
estimate the generative laws that determine how behavior evolves under
controlled environmental inputs.

An important implication of this perspective concerns the role of
experimental design. In classical psychometric testing, items are
selected primarily to maximize reliability or discriminative power. In a
metrological framework, perturbations must instead be designed to
maximize the identifiability of the behavioral system. Different
perturbation policies may reveal different aspects of the system's
dynamics, just as different experimental probes in physics can reveal
complementary properties of a physical system. The design of
perturbation environments thus becomes an integral part of the
measurement process rather than a secondary consideration.

Another key element of metrology is the notion of reproducibility across
instruments and laboratories. For behavioral measurement, this
requirement implies that reconstructed models should produce consistent
trajectory laws when behavioral systems are probed under equivalent
perturbation policies. Achieving such reproducibility will likely
require the development of standardized perturbation protocols, shared
experimental environments, and common data representations that allow
results to be compared across studies. Advances in immersive
technologies and digital experimental platforms make such
standardization increasingly feasible.

The metrological perspective also highlights the importance of
validation procedures that go beyond conventional goodness-of-fit
metrics. In the present framework, the criterion of functional
equivalence under unseen perturbations plays a central role. A
reconstructed model is considered an adequate representation of the
behavioral system if it generates trajectory distributions that match
those produced by the real system when exposed to novel perturbation
sequences. This requirement ensures that the reconstructed model
captures not only statistical regularities in observed data but also the
underlying dynamical principles governing behavioral responses.

Developing a full metrology of human behavior remains an ambitious goal.
It will require integrating advances from several disciplines, including
psychometrics, dynamical systems theory, computational modeling,
experimental design, and human--computer interaction. Nevertheless, the
perturbation-based reconstruction framework provides a conceptual
foundation for such an endeavor. By treating behavioral environments as
measurement instruments and by focusing on the reconstruction of
generative dynamics, it opens a path toward a more principled and
experimentally grounded science of behavioral measurement.

\section{Challenges}

The perturbation-based reconstruction framework outlined in this paper
suggests that behavioral measurement may be redefined as the
identification of dynamical systems through controlled interaction with
experimental environments. While this perspective opens new
methodological possibilities, it also raises a set of fundamental
challenges that must be addressed for a fully developed science of
behavioral measurement to emerge.

A first challenge concerns the design of perturbation environments
capable of efficiently probing behavioral systems. The informativeness
of reconstruction depends critically on the structure of the
perturbations used to interrogate the system. Designing perturbation
policies that maximize discriminability between competing models while
remaining ecologically meaningful and ethically acceptable remains an
open problem. This challenge becomes even more complex in adaptive
settings, where perturbations evolve in response to ongoing behavior and
must balance exploration and stability.

A second challenge relates to the standardization of behavioral
measurement procedures. In physical metrology, measurement depends on
well-defined instruments and protocols that ensure reproducibility
across laboratories. Achieving a comparable level of standardization in
behavioral science will require the development of shared perturbation
protocols, common experimental platforms, and interoperable data
representations. Programmable environments, including immersive and
VR-based systems, provide a potential infrastructure for such
standardization, but their widespread adoption raises issues of
comparability, calibration, and validation across different
implementations.

A third challenge concerns the identifiability and scalability of
behavioral system reconstruction. Human behavior is inherently
high-dimensional, stochastic, and context-dependent. Developing model
classes that are sufficiently expressive to capture this complexity
while remaining identifiable from finite data represents a major
methodological problem. In addition, reconstruction methods must scale
to large populations while preserving sensitivity to individual
differences in dynamical structure.

A fourth challenge involves the integration of multimodal behavioral
data. Advances in sensing technologies enable the simultaneous recording
of movement, physiology, gaze, and interaction patterns, generating
high-dimensional trajectory data. While such data increase
observability, they also introduce challenges related to data
integration, noise modeling, and computational complexity. Establishing
principled methods for combining multimodal signals into coherent
dynamical representations of behavior remains an open research
direction.

A fifth challenge concerns the role of artificial agents and generative
models in behavioral experimentation. The emergence of generative
artificial intelligence makes it possible to construct adaptive
environments in which perturbations are dynamically generated in
response to participant behavior. Such systems can potentially enhance
the efficiency of system identification by actively exploring
informative regions of the behavioral state space. However, they also
raise questions about interpretability, control, and the extent to which
artificially generated interactions faithfully represent real-world
behavioral contexts.

A sixth challenge relates to the definition of behavioral quantities and
units within a metrological framework. In the physical sciences,
measurement is grounded in well-defined units and invariant
transformations. In behavioral science, the corresponding quantities
remain less clearly defined. If behavioral systems are to be measured as
dynamical entities, it becomes necessary to identify which aspects of
their dynamics constitute meaningful and comparable quantities, such as
stability, flexibility, or resilience, and how these can be
operationalized across different experimental settings.

Finally, a broader conceptual challenge concerns the integration of this
framework within the existing landscape of behavioral science. The
perturbation-based perspective does not replace established approaches
such as psychometrics or computational modeling, but rather extends them
toward a more dynamical and experimentally grounded conception of
measurement. Developing a coherent synthesis that preserves the
strengths of existing methodologies while enabling system-level
reconstruction will be essential for the adoption of this paradigm.

Addressing these challenges will require coordinated advances in
experimental design, computational modeling, measurement theory, and
technological infrastructure. While substantial work remains, the
convergence of programmable environments, multimodal sensing, and
generative modeling suggests that the conditions necessary for a new
science of behavioral measurement are beginning to emerge.

\section{Discussion}\label{discussion}

The framework proposed in this paper suggests a shift in how behavioral
phenomena may be conceptualized and measured within the behavioral
sciences. Rather than treating behavior primarily as the manifestation
of static latent traits, the perturbation-based reconstruction
perspective conceptualizes behavior as the observable output of an
evolving dynamical system. Within this view, measurement consists not
only in estimating latent quantities but in reconstructing the
generative mechanisms that govern behavioral responses to controlled
environmental perturbations.

This perspective places behavioral science more closely with disciplines
in which system identification plays a central methodological role. In
fields such as physics, engineering, and control theory, systems are
studied by systematically probing them with controlled inputs and
analyzing the resulting trajectories. The goal is not merely to describe
observations but to infer the underlying dynamical structure capable of
generating those observations. The framework developed here extends this
logic to the study of human behavior by interpreting experimental
environments as instruments that interact with behavioral systems
through structured perturbations.

Historically, such an approach has been difficult to implement in the
behavioral sciences. Experimental paradigms have typically faced a
trade-off between ecological validity and experimental control.
Laboratory tasks provide precise manipulation and repeatability but
often capture only simplified aspects of behavior, while naturalistic
observations and experience sampling methods offer ecological realism
but limited control over the perturbations experienced by participants.
As a consequence, most measurement frameworks have focused on
aggregating responses into stable scores rather than reconstructing the
underlying behavioral dynamics that generate those responses.

Recent technological developments are beginning to change this
situation. Immersive experimental environments, particularly those
enabled by virtual reality, allow researchers to design perturbation
protocols that are both systematically controllable and ecologically
meaningful. Within such environments, the context, timing, and intensity
of perturbations can be precisely manipulated while maintaining a level
of experiential realism that approximates real-world situations. At the
same time, advances in sensing technologies make it possible to record
high-dimensional behavioral trajectories, including movement, gaze
patterns, physiological responses, and interaction dynamics.

These developments create the experimental conditions required for
perturbation-based reconstruction of behavioral systems. Virtual reality
environments can function as programmable measurement platforms in which
controlled environmental manipulations serve as probing signals, while
multimodal behavioral recordings provide the observable trajectories
from which the system's dynamics can be inferred. In this sense,
immersive environments play a role analogous to experimental apparatus
in the physical sciences: they enable systematic interactions between
the measurement instrument and the system under investigation.

The emergence of generative artificial intelligence further amplifies
these possibilities. Generative models and adaptive agents can be used
to construct interactive environments in which perturbations evolve
dynamically in response to the participant's behavior. This makes it
possible to design adaptive perturbation policies that explore the
behavioral system more efficiently, potentially improving the
identifiability of the reconstructed dynamics. Moreover, artificial
agents embedded in immersive environments can simulate complex social
interactions, enabling the systematic study of behavioral systems that
emerge from multi-agent dynamics.

Taken together, these technological and methodological developments
suggest that behavioral science may be entering a phase in which the
reconstruction of dynamical behavioral systems becomes experimentally
feasible. The perturbation-based framework proposed in this work
provides a conceptual structure for this transition by linking
controlled perturbations, behavioral trajectories, and generative
modeling within a unified measurement perspective.

Recent advances in generative and predictive modeling have demonstrated
an increasing ability to simulate realistic behavioral trajectories.
However, prediction alone does not constitute measurement. In the
absence of controlled perturbations and validation under novel
conditions, such models may capture statistical regularities in observed
data without identifying the underlying behavioral system. From this
perspective, the distinction between predictive adequacy and measurement
validity becomes central to the development of a metrology of human
behavior.

At the same time, several challenges remain before a full metrology of
human behavior can be realized. The design of perturbation policies that
efficiently probe behavioral systems remains an open problem.
Standardized experimental environments and perturbation protocols will
likely be required to ensure reproducibility across laboratories. In
addition, the development of model classes capable of capturing the
complexity of human behavior while remaining identifiable from empirical
data represents an important methodological challenge.

Despite these challenges, the perturbation-based reconstruction
framework highlights a promising direction for the future of behavioral
measurement. By treating experimental environments as instruments,
perturbations as measurement operations, and behavioral trajectories as
dynamical signals, it opens the possibility of moving beyond static
representations of behavior toward the identification of generative
behavioral systems. Such a shift could provide a foundation for a more
principled and experimentally grounded science of human behavior.

\section{Conclusions}

In this perspective, the measurement of human behavior may evolve from
the estimation of latent traits toward the identification of dynamical
behavioral systems. Rather than treating behavioral observations as
isolated responses to predefined stimuli, the perturbation-based
reconstruction framework interprets them as trajectories generated by an
underlying system interacting with its environment. Measurement thus
becomes an active process in which experimental environments are
designed to probe the structure of behavioral dynamics through
controlled perturbations.

Advances in immersive environments, sensing technologies, and
computational modeling now make it possible to implement this approach
experimentally. Virtual reality environments can function as
programmable measurement platforms in which complex yet controllable
perturbations are applied to behavioral systems, while multimodal
sensing allows the continuous observation of behavioral trajectories.
When combined with generative models and adaptive perturbation
strategies, these environments enable a form of experimental probing
that was previously unattainable in behavioral research.

The framework proposed here therefore points toward the possibility of a
more principled approach to behavioral measurement grounded in system
identification. By integrating controlled perturbations, dynamical
modeling, and trajectory-level validation, it provides a conceptual
foundation for reconstructing the generative mechanisms underlying human
behavior. Although substantial methodological challenges remain, the
convergence of immersive technologies, computational modeling, and
experimental design suggests that the reconstruction of behavioral
dynamics may become a central methodological paradigm for future
behavioral science. More broadly, this perspective reframes behavioral
measurement as the problem of identifying generative systems through
controlled interactions, bringing it conceptually closer to the
measurement traditions of the physical sciences while preserving the
complexity of human behavior.

In this sense, the long-standing ambition of developing a rigorous
measurement science of human behavior, analogous in spirit to metrology
in the physical sciences, may begin to move from conceptual aspiration
toward practical possibility.

\appendix
\section{Simulation Procedure}

The following pseudo-code illustrates the core steps of the simulation
and reconstruction procedure used in the study. Full implementation
details are provided in the Supplementary Materials (\textbf{R code}).

\textit{Perturbation-based behavioral system reconstruction (simulation)}

Input: System parameters (A, B, C), noise models, perturbation regimes\\
Output: Reconstructed model $\hat{M}$ and discrepancy measures

\begin{enumerate}
\def\labelenumi{\arabic{enumi}.}
\item
  Define ground-truth dynamical system\\
  $x_{t+1} = A x_t + B u_t + w_t$\\
  $y_t = C x_t + v_t$
\item
  Generate perturbation sequences\\
  a. Rich regime: $u_t \sim$ distribution with sufficient variability\\
  b. Poor regime: $u_t \approx$ constant or low-variance process
\item
  Simulate trajectory dataset\\
  For each trajectory i = 1,...,N:\\
  Generate $(x_t^{(i)}, y_t^{(i)})$ under chosen perturbation
  regime
\item
  Estimate model from data\\
  Fit parameters ($\hat{A}$, $\hat{B}$) using system identification on training
  trajectories
\item
  Evaluate on unseen perturbations\\
  Generate new input sequence $u_{\mathrm{test}}$\\
  Simulate trajectories from both true system and reconstructed model
\item
  Assess functional equivalence\\
  Compute discrepancy D between trajectory distributions\\
  (e.g., Wasserstein distance)
\item
  Compare regimes\\
  Analyze D under rich vs poor perturbations
\end{enumerate}

Return: Estimated model and discrepancy measures

\section*{References}

Åström, K. J. (1995). Adaptive control. In Mathematical System Theory:
The Influence of RE Kalman (pp. 437-450). Berlin, Heidelberg: Springer
Berlin Heidelberg.

Åström, K. J., \& Eykhoff, P. (1971). System identification-a survey.
Automatica, 7(2), 123-162.

Bellman, R., \& Åström, K. J. (1970). On structural identifiability.
Mathematical biosciences, 7(3-4), 329-339.

Bohil CJ, Alicea B, Biocca FA. Virtual reality in neuroscience research
and therapy. Nat Rev Neurosci. 2011 Nov 3;12(12):752-62. doi:
10.1038/nrn3122. PMID: 22048061.

Borel, A. (2012). Linear algebraic groups. Springer Science \& Business
Media.

Borsboom, D. (2005). Measuring the mind: Conceptual issues in
contemporary psychometrics. Cambridge University Press.

Box, G. E. P. (1976). Science and Statistics. Journal of the American
Statistical Association, 71(356), 791--799.
https://doi.org/10.1080/01621459.1976.10480949

Bronfenbrenner, U. (1977). Toward an experimental ecology of human
development. American Psychologist, 32(7), 513--531.
https://doi.org/10.1037/0003-066X.32.7.513

Chirico, A., Yaden, D. B., Riva, G., \& Gaggioli, A. (2016). The
potential of virtual reality for the investigation of awe. Frontiers in
psychology, 7, 1766.

Cipresso P (2015) Modeling behavior dynamics using computational
psychometrics within virtual worlds. Front. Psychol. 6:1725. doi:
10.3389/fpsyg.2015.01725

Cipresso P and Immekus JC (2017) Back to the Future of Quantitative
Psychology and Measurement: Psychometrics in the Twenty-First Century.
Front. Psychol. 8:2099. doi: 10.3389/fpsyg.2017.02099

Cipresso, P., Giglioli, I. A. C., Raya, M. A., \& Riva, G. (2018). The
past, present, and future of virtual and augmented reality research: a
network and cluster analysis of the literature. Frontiers in psychology,
9, 2086.

Cover, T. M., \& Thomas, J. A. (2012). Elements of Information Theory.
John Wiley \& Sons.

Cronbach, L. J., \& Meehl, P. E. (1955). Construct validity in
psychological tests. Psychological Bulletin, 52(4), 281--302.
https://doi.org/10.1037/h0040957

Csikszentmihalyi, M., Larson, R. (2014). Validity and Reliability of the
Experience-Sampling Method. In: Flow and the Foundations of Positive
Psychology. Springer, Dordrecht.
https://doi.org/10.1007/978-94-017-9088-8\_3

Durbin, J., \& Koopman, S. J. (2012). Time series analysis by state
space methods. Oxford University Press (UK).

Eddy, S. R. (2004). What is a hidden Markov model?. Nature
biotechnology, 22(10), 1315-1316.

Faria AL, Latorre J, Silva Cameirão M, Bermúdez i Badia S and Llorens R
(2023) Ecologically valid virtual reality-based technologies for
assessment and rehabilitation of acquired brain injury: a systematic
review. Front. Psychol. 14:1233346. doi: 10.3389/fpsyg.2023.1233346

Friston, K. (2010). The free-energy principle: a unified brain theory?.
Nature reviews neuroscience, 11(2), 127-138.

Fritz, J., Piccirillo, M. L., Cohen, Z. D., Frumkin, M., Kirtley, O.,
Moeller, J., Neubauer, A. B., Norris, L. A., Schuurman, N. K., Snippe,
E., \& Bringmann, L. F. (2024). So you want to do ESM? 10 essential
topics for implementing the experience-sampling method. Advances in
Methods and Practices in Psychological Science, 7(3), Article
25152459241267912. https://doi.org/10.1177/25152459241267912

Gevers, M., \& Ljung, L. (1986). Optimal experiment designs with respect
to the intended model application. Automatica, 22(5), 543-554.

Ghaffarzadegan, N., Majumdar, A., Williams, R., \& Hosseinichimeh, N.
(2024). Generative agent‐based modeling: an introduction and tutorial.
System Dynamics Review, 40(1), e1761.

Ghahramani, Z., \& Hinton, G. E. (1996). Switching state-space models.
University of Toronto Technical Report CRG-TR-96-3, Department of
Computer Science.

Gideon Schwarz. "Estimating the Dimension of a Model." Ann. Statist. 6
(2) 461 - 464, March, 1978. https://doi.org/10.1214/aos/1176344136

Gilbert, N. (2019). Agent-based models. Sage Publications.

Goodfellow, I., Pouget-Abadie, J., Mirza, M., Xu, B., Warde-Farley, D.,
Ozair, S., ... \& Bengio, Y. (2020). Generative adversarial networks.
Communications of the ACM, 63(11), 139-144.

Gretton, A., Borgwardt, K. M., Rasch, M. J., Schölkopf, B., \& Smola, A.
(2012). A kernel two-sample test. The journal of machine learning
research, 13(1), 723-773.

Hamaker EL, Asparouhov T, Brose A, Schmiedek F, Muthén B. At the
Frontiers of Modeling Intensive Longitudinal Data: Dynamic Structural
Equation Models for the Affective Measurements from the COGITO Study.
Multivariate Behav Res. 2018 Nov-Dec;53(6):820-841. doi:
10.1080/00273171.2018.1446819. Epub 2018 Apr 6. PMID: 29624092.

Hamaker, E. L., Ceulemans, E., Grasman, R. P., \& Tuerlinckx, F. (2015).
Modeling affect dynamics: State of the art and future challenges.
Emotion Review, 7(4), 316-322.

Hamaker, E. L., \& Wichers, M. (2017). No time like the present:
Discovering the hidden dynamics in intensive longitudinal data. Current
Directions in Psychological Science, 26(1), 10--15.
https://doi.org/10.1177/0963721416666518

Hollenstein, T., \& Lewis, M. D. (2006). A state space analysis of
emotion and flexibility in parent-child interactions. Emotion, 6(4),
656.

Kahneman, D., Krueger, A. B., Schkade, D. A., Schwarz, N., \& Stone, A.
A. (2004). A survey method for characterizing daily life experience: The
day reconstruction method. Science, 306(5702), 1776-1780.

Kelso, J. S. (1995). Dynamic patterns: The self-organization of brain
and behavior. MIT press.

Kitagawa, G., \& Gersch, W. (1996). Linear Gaussian state space
modeling. In Smoothness priors analysis of time series (pp. 55-65). New
York, NY: Springer New York.

Kloeden, P. E., \& Pearson, R. A. (1977). The numerical solution of
stochastic differential equations. The ANZIAM Journal, 20(1), 8-12.

Kullback, S., \& Leibler, R. A. (1951). On information and sufficiency.
The annals of mathematical statistics, 22(1), 79-86.

Kuppens, P., Oravecz, Z., \& Tuerlinckx, F. (2010). Feelings change:
Accounting for individual differences in the temporal dynamics of
affect. Journal of Personality and Social Psychology, 99(6), 1042--1060.
https://doi.org/10.1037/a0020962

Ljung, L., ``Perspectives on system identification,'' Annual Reviews in
Control, vol. 34, pp. 1-12, 2010.

Markov, K., Matsui, T., Septier, F., \& Peters, G. (2015, August).
Dynamic speech emotion recognition with state-space models. In 2015 23rd
European Signal Processing Conference (EUSIPCO) (pp. 2077-2081). IEEE.

Marsman, M., Waldorp, L., \& Borsboom, D. (2023). Towards an
encompassing theory of network models: Reply to Brusco, Steinley,
Hoffman, Davis-Stober, and Wasserman (2019).~\emph{Psychological
Methods, 28}(4), 757--764.~https://doi.org/10.1037/met0000373

Molenaar, P. C. M. (2004). A Manifesto on Psychology as Idiographic
Science: Bringing the Person Back Into Scientific Psychology, This Time
Forever. Measurement: Interdisciplinary Research and Perspectives, 2(4),
201--218. https://doi.org/10.1207/s15366359mea0204\_1

Muandet, K., Fukumizu, K., Sriperumbudur, B., \& Schölkopf. B. (2017).
Kernel mean embedding of distributions: A review and beyond. Foundations
and Trends in Machine Learning, 10(1-2), 1-141.

Oksendal, B. (2013). Stochastic differential equations: an introduction
with applications. Springer Science \& Business Media.

Pavliotis, G. A. (2014). Stochastic processes and applications. Texts in
applied mathematics, 60, 41-43.

Peyré, G., \& Cuturi, M. (2019). Computational optimal transport: With
applications to data science. Now Foundations and Trends.

Platen, E., \& Bruti-Liberati, N. (2010). Numerical solution of
stochastic differential equations with jumps in finance (Vol. 64).
Springer Science \& Business Media.

Pooja R, Ghosh P, Sreekumar V. Towards an ecologically valid
naturalistic cognitive neuroscience of memory and event cognition.
Neuropsychologia. 2024 Oct 10;203:108970. doi:
10.1016/j.neuropsychologia.2024.108970. Epub 2024 Aug 13. PMID:
39147361.

Potter LN, Yap J, Dempsey W, Wetter DW, Nahum-Shani I. Integrating
Intensive Longitudinal Data (ILD) to Inform the Development of Dynamic
Theories of Behavior Change and Intervention Design: a Case Study of
Scientific and Practical Considerations. Prev Sci. 2023
Nov;24(8):1659-1671. doi: 10.1007/s11121-023-01495-4. Epub 2023 Apr 15.
PMID: 37060480; PMCID: PMC10576833.

Riva, G. (2009). Virtual reality: an experiential tool for clinical
psychology. British Journal of Guidance \& Counselling, 37(3), 337-345.

Riva, G., Wiederhold, B. K., \& Mantovani, F. (2019). Neuroscience of
virtual reality: from virtual exposure to embodied medicine.
Cyberpsychology, behavior, and social networking, 22(1), 82-96.

Schölkopf, B., \& Smola, A. J. (2002). Learning with kernels: support
vector machines, regularization, optimization, and beyond. MIT press.

Shiffman S, Stone AA, Hufford MR. Ecological momentary assessment. Annu
Rev Clin Psychol. 2008;4:1-32. doi:
10.1146/annurev.clinpsy.3.022806.091415. PMID: 18509902.

Santambrogio, F. (2015). Wasserstein distances and curves in the
Wasserstein spaces. In: Optimal Transport for Applied Mathematicians.
Progress in Nonlinear Differential Equations and Their Applications, vol
87. Birkhäuser, Cham. https://doi.org/10.1007/978-3-319-20828-2\_5

Slater M and Sanchez-Vives MV (2016) Enhancing Our Lives with Immersive
Virtual Reality. Front. Robot. AI 3:74. doi: 10.3389/frobt.2016.00074

Söderström, T., \& Stoica, P. (Eds.). (1988). System identification.
Prentice-Hall, Inc.

Speyer, L. G., Murray, A. L., \& Kievit, R. (2024). Investigating
Moderation Effects at the Within-Person Level Using Intensive
Longitudinal Data: A Two-Level Dynamic Structural Equation Modelling
Approach in Mplus. Multivariate Behavioral Research, 59(3), 620--637.
https://doi.org/10.1080/00273171.2023.2288575

Strogatz, S. H. (2024). Nonlinear dynamics and chaos: with applications
to physics, biology, chemistry, and engineering. Chapman and Hall/CRC.

Thelen, E., \& Smith, L. B. (1994). A dynamic systems approach to the
development of cognition and action. MIT press.

Valenza, G., Lanata, A., \& Scilingo, E. P. (2011). The role of
nonlinear dynamics in affective valence and arousal recognition. IEEE
transactions on affective computing, 3(2), 237-249.

Van Gelder, T. (1995). What might cognition be, if not computation?. The
journal of Philosophy, 92(7), 345-381.

Villani, C. (2009). The Wasserstein distances. In: Optimal Transport.
Grundlehren der mathematischen Wissenschaften, vol 338. Springer,
Berlin, Heidelberg. https://doi.org/10.1007/978-3-540-71050-9\_6

Waugh, C. E., \& Kuppens, P. (Eds.). (2021). Affect dynamics. Springer
Nature.

Wiggins, S. (2003). Introduction to applied nonlinear dynamical systems
and chaos. New York, NY: Springer New York.

Xie, K., Vongkulluksn, V. W., Heddy, B. C., \& Jiang, Z. (2024).
Experience sampling methodology and technology: An approach for
examining situational, longitudinal, and multi-dimensional
characteristics of engagement. Educational Technology Research and
Development, 72(5), 2585--2615.
https://doi.org/10.1007/s11423-023-10259-4

\end{document}